\documentclass[useAMS,usenatbib]{mn2e}
\usepackage{times}
\usepackage{graphicx}
\usepackage{amssymb}
\usepackage{natbib}
\usepackage{psfig}
\title[A Quantitative Link Between Globular Clusters and the Stellar Halos in Elliptical Galaxies]{A Quantitative Link Between Globular Clusters and the Stellar Halos in Elliptical Galaxies}
\author[Forte et al.]{Juan C. Forte$^{1,2}$\thanks{E-mail:forte@fcaglp.unlp.edu.ar}, Favio Faifer$^{1,2,3}$\thanks{E-mail:favio@fcaglp.unlp.edu.ar} and Doug Geisler $^{4}$\thanks{E-mail:dgeisler@astro.udec.cl}\thanks{Visiting Astronomer Cerro Tololo Interamerican Observatorty and Kitt Peak National Observatory,  operated by AURA under contract with the National Science Foundation}\\
$^1${Facultad de Ciencias Astron\'omicas y Geof\'isicas, Universidad Nacional de La Plata, Paseo del Bosque s/n, 1900 La Plata, Argentina}\\
$^2${Consejo Nacional de Investigaciones Cient\'ificas y T\'ecnicas, Rep.
Argentina}\\
$^3${IALP}\\
$^4${Grupo de Astronom\'\i a, Departamento de F\'\i sica, Casilla 160, Universidad de Concepci\'on, Chile}\\
}
\begin{document}

\date{Accepted 2007 September 22. Received 2007 September 13; in original form 2007 August 10 }
\pagerange{\pageref{firstpage}--\pageref{lastpage}} \pubyear{}
\maketitle

\label{firstpage}

\begin{abstract}

  This paper explores the quantitative connection between globular
  clusters  and the ``diffuse'' stellar population of the galaxies they
  are associated with. Both NGC 1399 and NGC 4486 (M87)  are well
  suited for this kind of analysis due to their large globular cluster
  populations. 

  The main assumption of our Monte Carlo based models is that each globular
  cluster is formed   along with a given diffuse stellar mass that
  shares the same spatial distribution, chemical composition and age. The main
  globular clusters subpopulations, that determine the observed bimodal colour
  distribution, are decomposed avoiding {\it a priori} parametric
  (e.g. Gaussian) fits and using a new colour (C-T$_1$)-metallicity
  relation. The eventual detectability
  of a ``blue'' tilt in the colour magnitude diagrams of the blue globulars
  subpopulation is also addressed.
  
  A successful link between globular clusters and the stellar galaxy halo 
  is established by assuming that the number of globular 
  clusters per associated diffuse stellar mass t is a function of total
  abundance [Z/H] and behaves as $t=\gamma \exp(-\delta[Z/H])$ (i.e. increases
  when abundance decreases).  

  The simulations allow the prediction of a surface brightness
  profile for each galaxy through this two free parameters approximation.
  The $\gamma$, $\delta$  parameters that provide the best fit to the observed 
  profiles in the B band, in turn, determine several features, namely, large
  scale halo colour gradients, globular clusters-halo colour offset, clusters
  cumulative specific frequencies, and  stellar metallicity distributions,
  that compare well with observations. 

  The results suggest the coexistence of two distinct stellar populations
  characterised by widely different metallicities and spatial distributions.
  One of these populations (connected with the blue globulars) is metal poor, 
  highly homogeneous, exhibits an extended spatial distribution and becomes
  more evident at large galactocentric radius contributing with some 
  20\% of the total stellar mass. In turn, the stellar population
  associated with the red globulars is extremely heterogeneous and
  dominates the inner region of both galaxies.  

  Remarkably, and although the cluster populations of these galaxies exhibit
  detectable differences in colour distribution, the $\delta$ parameter that 
  determines the shape  of the brightness profiles of both galaxies has the
  same value, $\delta \approx$ 1.1 to 1.2 $\pm 0.1$.

\end{abstract}

\begin{keywords}
galaxies: star clusters: general -- galaxies: globular clusters: individual: NGC 1399, NGC 4486 -- galaxies:haloes
\end{keywords}
%=====================================================================
\section{Introduction}
\label{INTRO}

  The idea that globular clusters (GCs) harbour important clues
  in relation with the early stages of galaxy formation is a widely
  accepted concept. One of the most compelling arguments in favour
  of the existence of a connection between GCs and major star formation
  episodes  in the life of a galaxy is the constant cluster formation
  efficiency, defined in terms of total baryonic mass \citep{b46},
  in different galaxies.

  However, breaking the code that leads to a detailed  quantitative link 
  between GCs and the underlying ``diffuse'' stellar population is still 
  an open question. Such a connection has been discussed on theoretical
   (e.g. \citealt*{b3} or, more recently, \citealt{b56b}) and observational
   grounds (e.g. \citealt{b20}). In the particular case of Milky Way GCs, \citet*{b57b}
   found that the chemical similarities between clusters and field stars with
 $[Fe/H]\le -1$ suggests a shared chemical history in a well mixed early Galaxy.
 
  Clarifying this issue may certainly yield some arguments in favour
  (or against) some predominant ideas that have been widely referenced
  in the literature (e.g. \citealt{b16}; \citealt{b72b})
  and later explored within the frame of different
  scenarios (e.g. \citealt{b1}, or \citealt{b18}).
 
  A good perspective of the complex situation in this context is given
  in the thorough review by \citet{b5} and \citet{b41}.
                                                                        
  An initial confrontation between GCs and halo stellar populations
  shows more differences than similarities: a) In general, GCs 
  exhibit more shallow spatial distributions than those characterising
  galaxy light (e.g. \citealt{b59}; \citealt{b14}); b) There is a colour
  offset in the sense that mean integrated globular colours appear bluer
  than those of the galaxy halos at the same galactocentric radius
  (\citealt{b78}; \citealt*{b22}; \citealt{b37b}); c) GCs show
  frequently  bimodal colour (and hence, metallicity) distributions
  (see, for example, \citealt{b53}). This feature
  does not seem exactly shared by the resolved stellar populations  
  in nearby resolved galaxies (see \citealt{b15}; \citealt{b33}; \citealt{b62} or \citealt{b50}). 

  As discussed later in this work, those differences arise, mainly,
  from the fact that GCs analysis usually provide {\bf number} weighted statistics
  while galaxy halos observations yield  {\bf luminosity} weighted measurements.

  A preliminary quantitative approach to the globulars-stellar halo
  connection was presented in \citet*{b25} (hereafter
  FFG05). This last paper shows that, given the areal density 
  distribution of the ``blue'' and ``red'' globular cluster subpopulations 
  in NGC 1339, the galaxy   surface brightness profile, galactocentric
  colour gradient and  cumulative GCs specific frequency,
  can be matched by linearly weighting the areal density profiles. 
  The ``weight'' corresponding to each component of the brightness profile
  is the inverse of the {\bf intrinsic} GCs frequency characteristic of each 
  cluster population.

  The main argument behind that approach is that the shape of the colour 
  (and metallicity) distribution of each globular cluster subpopulation
  does not change with galactocentric radius. Large angular scale
  studies (\citealt{b14}; \citealt{b2}) in fact show that
  the colour peaks in the GCs colour statistics of NGC 1399
  keep the same position (or show very little variation) over large 
  galactocentric ranges. A similar result is obtained for NGC 4486 by
  \citet{b40} who find those peaks do not show a detectable
  variation in colour over 75 kpc in galactocentric radius.

  It must be stressed that those subpopulations are ``phenomenologically''
  defined in terms of their integrated colours but each might eventually 
  have  a given spread in age and/or metallicity.  
  
  The presence of a ``valley'' in the globular colour statistics is usually
  adopted as a discriminating boundary between both subpopulations. The need
  to revise such a procedure was already suggested by figure 5 in FFG05.  
  This last diagram showed that the NGC 1399 GCs bluer than the blue
  peak have a distinct behaviour of the areal density profile, exhibiting
  a flat core that disappears when all ``blue'' clusters (i.e. all GCs bluer
  than
  the colour valley) are included in the sample. That result prompted
  for a further analysis as discussed below.

  This paper generalises the FFG05 approach trough  Monte-Carlo based
  models. In this frame, ``seed'' globulars are generated following a given 
  abundance Z distribution and then associated with a ``diffuse'' stellar
  mass that shares its age, chemical composition and statistical spatial distribution. The
  luminosity associated with this mass is derived from a mass to luminosity
  ratio adequate for a given age and metallicity. These models aim at 
  reproducing the features mentioned above and seek for a function that
  could link GCs and diffuse stellar populations keeping a
  minimum number of free parameters. 

  Both NGC 1399 and NGC 4486 appear as adequate targets in order to
  perform such a modelling due to their prominent globular cluster
  systems (GCS). Although these
  systems show some structural similarities in terms of their spatial 
  distribution, they also differ markedly both in the shape of
  their GCs colour statistics  and specific frequencies \citep{b24}.

  This work also presents new Washington photometry, obtained and 
  handled in a homogeneous way, that allows for a re-discussion of
  the GCS properties in the inner region of both galaxies and, in 
  particular, of the behaviour of the areal density of GCs
  with colour. In turn, recent wide field photometric studies of the GCS 
  associated with NGC 1399 \citep{b2} and NGC 4486 (\citealt{b80}; \citealt{b81}), are well suited for extending the analysis to larger 
  galactocentric radii.

 \section{Observations and data handling}
 \label{ODH}

\begin{figure*}
\resizebox{0.4\hsize}{!}{\includegraphics{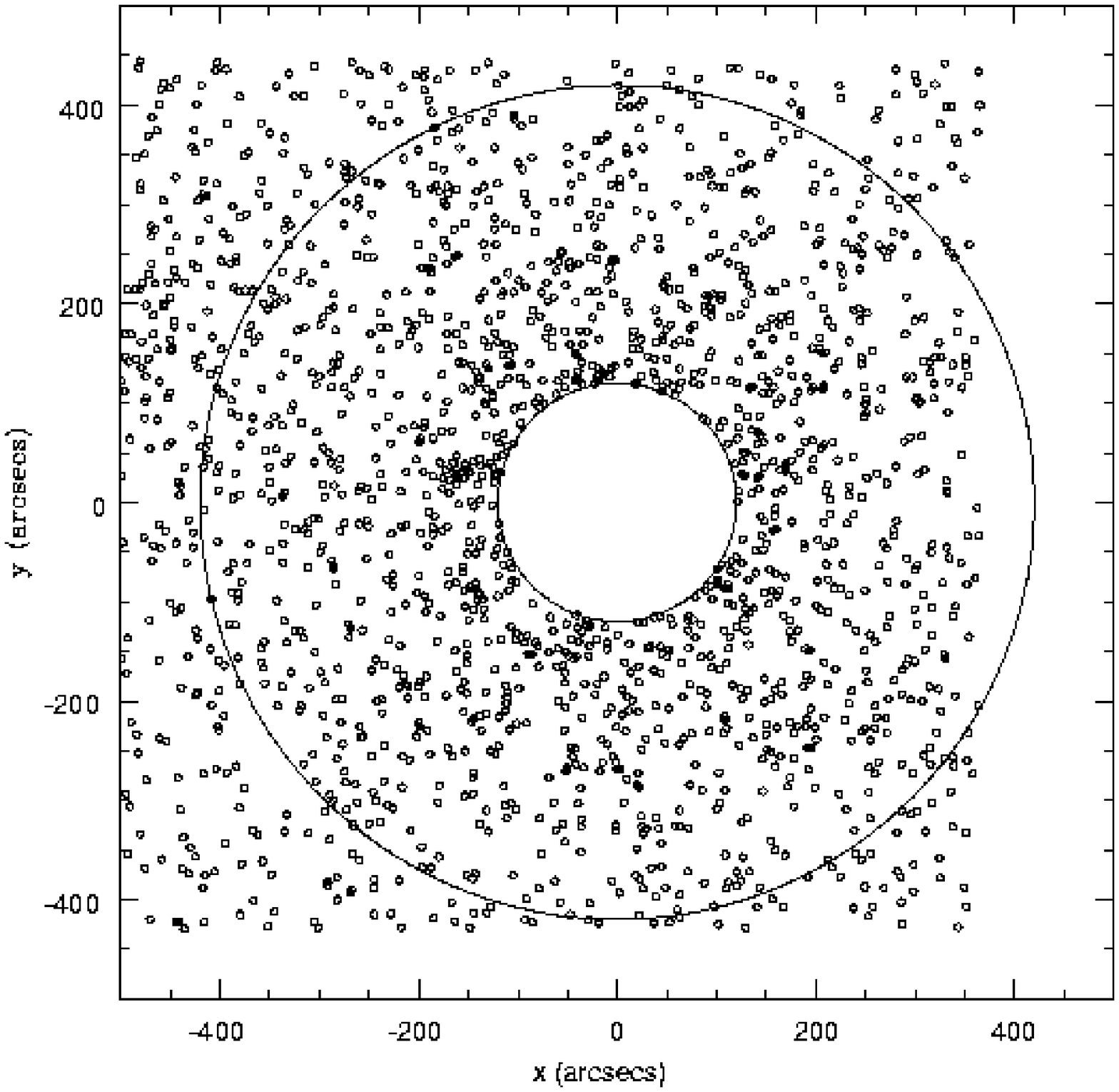}}
\resizebox{0.4\hsize}{!}{\includegraphics{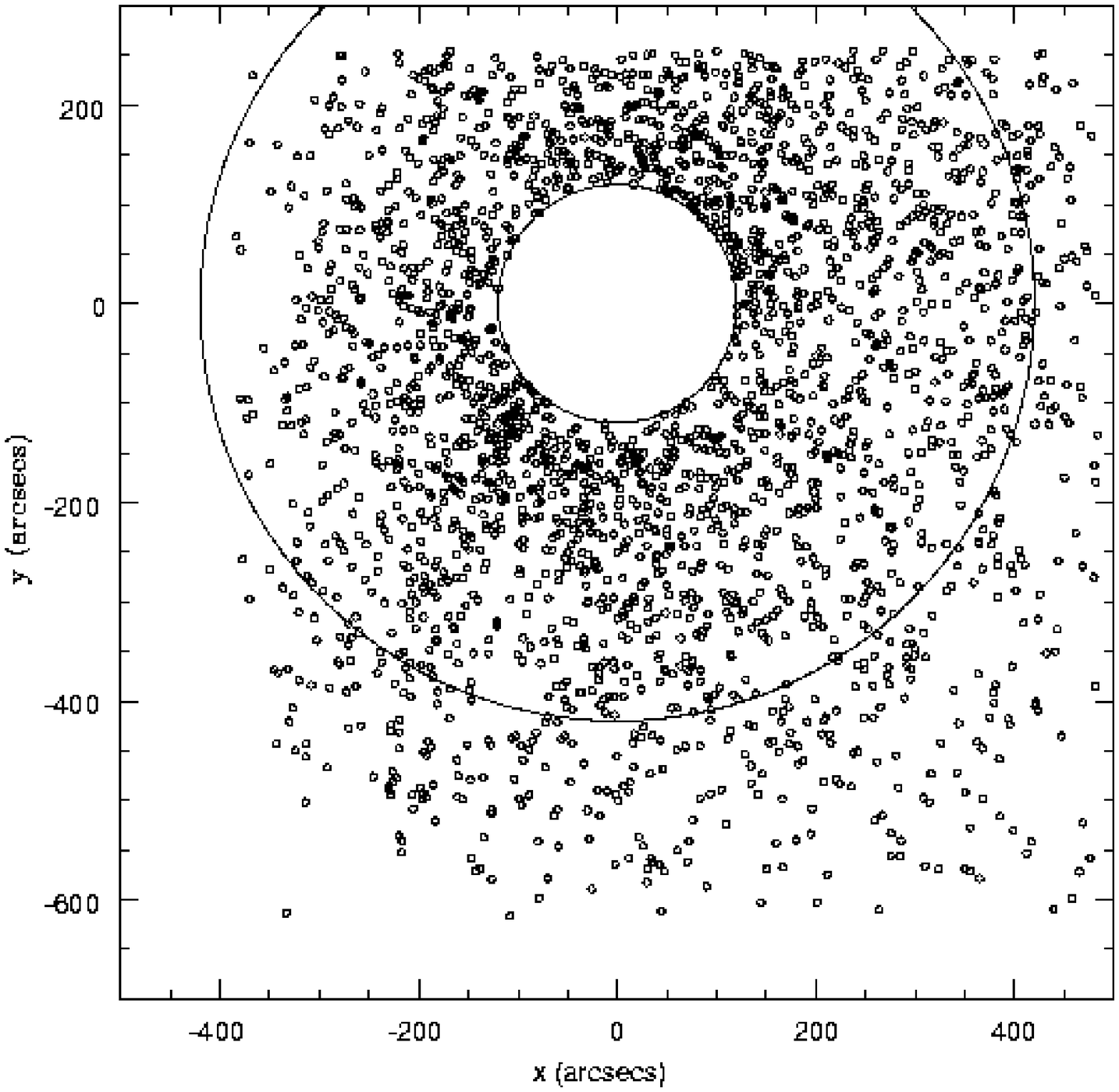}}
\caption{Distribution GCs candidates brighter than T$_1=23.2$. Left panel: NGC 1399 field. North is to the right and East is up. Right panel: NGC 4486 field. North is up and East to the left. In both panels, circles have radii of 120 and 420 arcsec respectively. The GC sample completeness  in these areas is close to 95\%. }
\label{Sample_sky}
\end{figure*}

  Photometric observations were carried out with the Mayall and Blanco
  4-m telescopes at KPNO and CTIO respectively and 2048 pixels on a
  side CCDs, with a pixel scale of 0.43 arcsecs. The C  filter from
  the Washington System \citep{b30} and the R$_{KC}$
  filter of the Kron-Cousins system were used at both telescopes. As 
  noted by \citet{b27} this last filter is comparable to the T$_1$ 
  filter in the Washington system although much more efficient in terms of 
  transmission. In what follows we keep the T$_1$ denomination for our
  red magnitudes.

  Two R$_{KC}$ images (exposure: 600 secs each) and three C
  images (exposure: 1500 secs each) were secured for both galaxy
  fields.

  Seeing quality for these frames varied between 1.0 and 1.6 arcsec.
  These images were processed with the CCDRED routines within IRAF 
  \footnote{IRAF is distributed by the National Optical
  Astronomical Observatories,which are operated by the Association of
  Universities for Research in Astronomy, Inc., under cooperative
  agreement with the National Science Foundation} 
  including bias and flat fielding.

 The galaxy background removal was performed with routins included in the VISTA image
 prossesing systems. 

  PSF photometry on all the frames was then carried out using the ALLFRAME version of
  the DAOPHOT package (\citealt{b74}; \citealt{b75}).

  The instrumental photometry was transformed to the standard system
  using  calibration standard stars from \citet{b27}.

  Image classification, in terms of resolved and non-resolved objects,
  was performed as in \citet{b23}. Briefly, that procedure
  combined the use of the round and sharpness parameters defined in 
  DAOPHOT and also the mirrored envelope of the T$_1$ PSF ALLFRAME
  magnitude vs. the difference between this magnitude and that obtained
  using aperture photometry for every object detected on the images.

  Non-resolved objects brighter than T$_1=23.2$ and with (C-T$_1$) colours
  between 0.9 and 2.3 were considered as cluster candidates and their
  distribution on the sky is depicted in Figure \ref{Sample_sky}.

  Circles with r=120 and 420 arcsec delineate the area used for the 
  analysis of the surface density distributions in the inner regions
  of the galaxies.

\begin{figure}
\resizebox{1.\hsize}{!}{\includegraphics{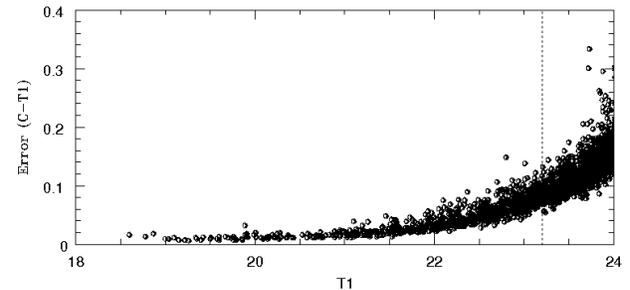}}
\caption{ (C-T$_1$) colour errors as a function of T$_1$ magnitude for both galaxy fields. The  vertical line at T$_1=23.2$ is the limiting magnitude adopted in the analysis.  The median colour error for the sample is 0.04 mags.}
\label{P_error}
\end{figure}

  Figure \ref{P_error} shows the errors on the (C-T$_1$) colours as a
  function of T$_1$
  magnitude as delivered by DAOPHOT. A median error for the (C-T$_1$) 
  colours of $\pm$ 0.07 mags. is reached at T$_1=23.2$, which is adopted in
  what follows as the limiting magnitude of the analysis in order to 
  assure good quality colours.

  The photometric data for both galaxies, Table \ref{Phot_data1399} and \ref{Phot_data4486},
 are available in  the electronic journal version. Coordinates are referred to the galaxy centers and defined as in Figure \ref{Sample_sky}

\begin{table}
\centering
%\begin{minipage}
\caption{Photometric data NGC 1399.}
\label{Phot_data1399}
\begin{tabular}{@{}cccccc@{}}
\hline
ID & X (arcsec)   &  Y (arcsec) & T$_1$ & (C-T$_1$) & roundness \\
\hline
          537.  &   196.8  &  -427.6  &   24.15 &    0.63  &   0.77\\
          553.  &   136.5  &  -427.5  &  24.33  &   0.65   &  0.85\\
\end{tabular}
%\end{minipage}
\end{table}

\begin{table}
\centering
%\begin{minipage}
\caption{Photometric data NGC 4486.}
\label{Phot_data4486}
\begin{tabular}{@{}cccccc@{}}
\hline
ID & X (arcsec)   &  Y (arcsec) & T$_1$ & (C-T$_1$) & roundness \\
\hline
        23314. &    -13.1  &  -126.5  &  21.67 &    1.43 &    0.93\\
        23406.  &   -32.1 &   -124.9  &  23.39  &   1.26  &   0.94\\
\end{tabular}
%\end{minipage}
\end{table}

  ADDSTARS experiments were carried out to estimate the completeness of
  the non-resolved objects (which is expected to be the case for
  GCs at the distances of NGC 1399 and NGC 4486 from the sun). Ten
  images, adding 1000 artificial stars each, on both C and T$_1$ master
  images, yielded a completeness of 94 and 96\% at T$_1=23.2$, for
  NGC 1399 and NGC 4486 respectively.

  A comparison field of 77.7 arcmin$^{\sq}$ was taken from \citet{b23}
  who performed C and T$_1$ photometry following the same procedure. 
  This field has 146 non-resolved objetcs within the colour-magnitude 
  boundaries adopted for the globular cluster candidates.

 \section{Colour-magnitude diagrams and Colour Distributions}
 \label{CMDCD}

\begin{figure}
\resizebox{1.0\hsize}{!}{\includegraphics{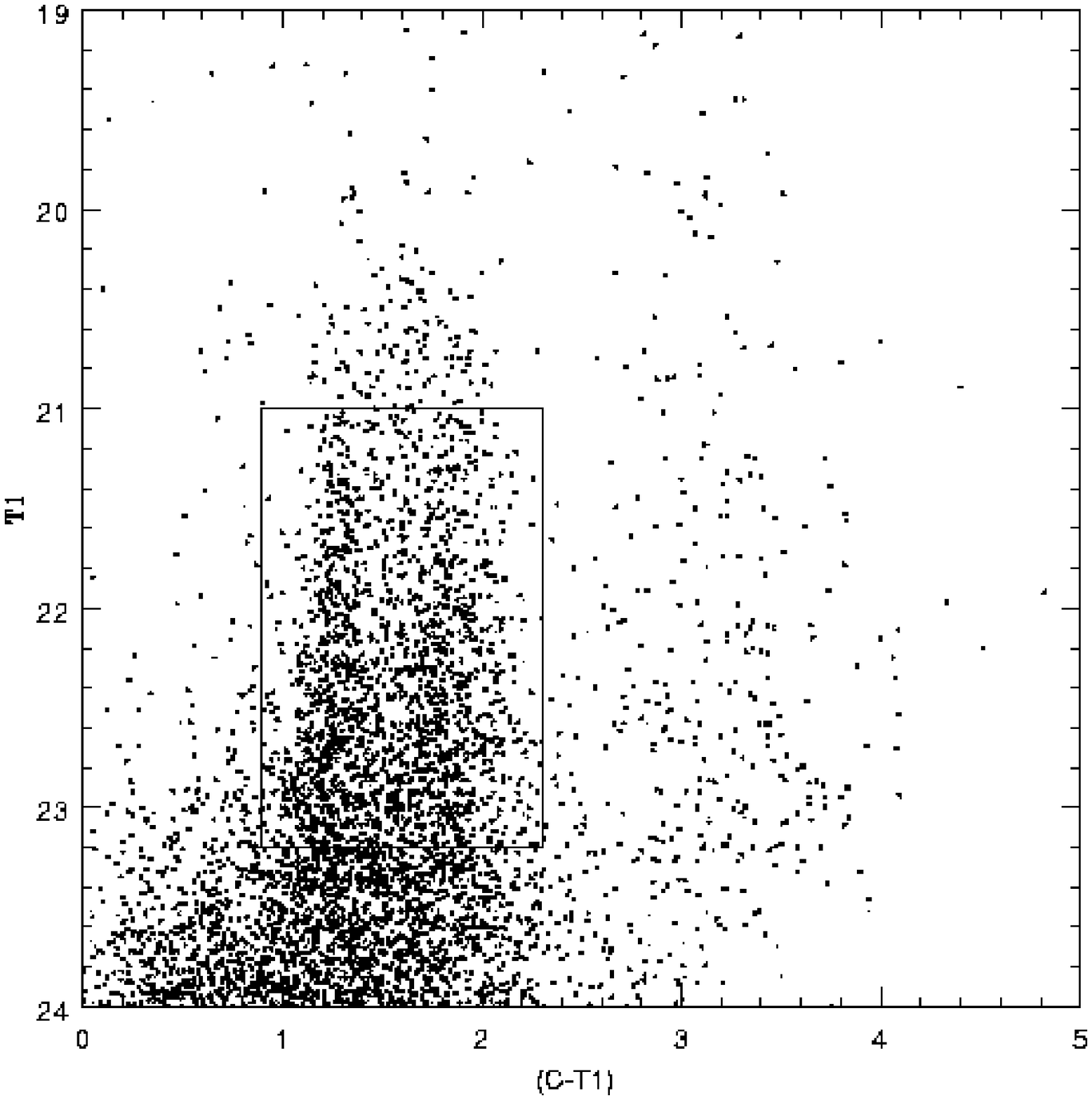}}
\resizebox{1.0\hsize}{!}{\includegraphics{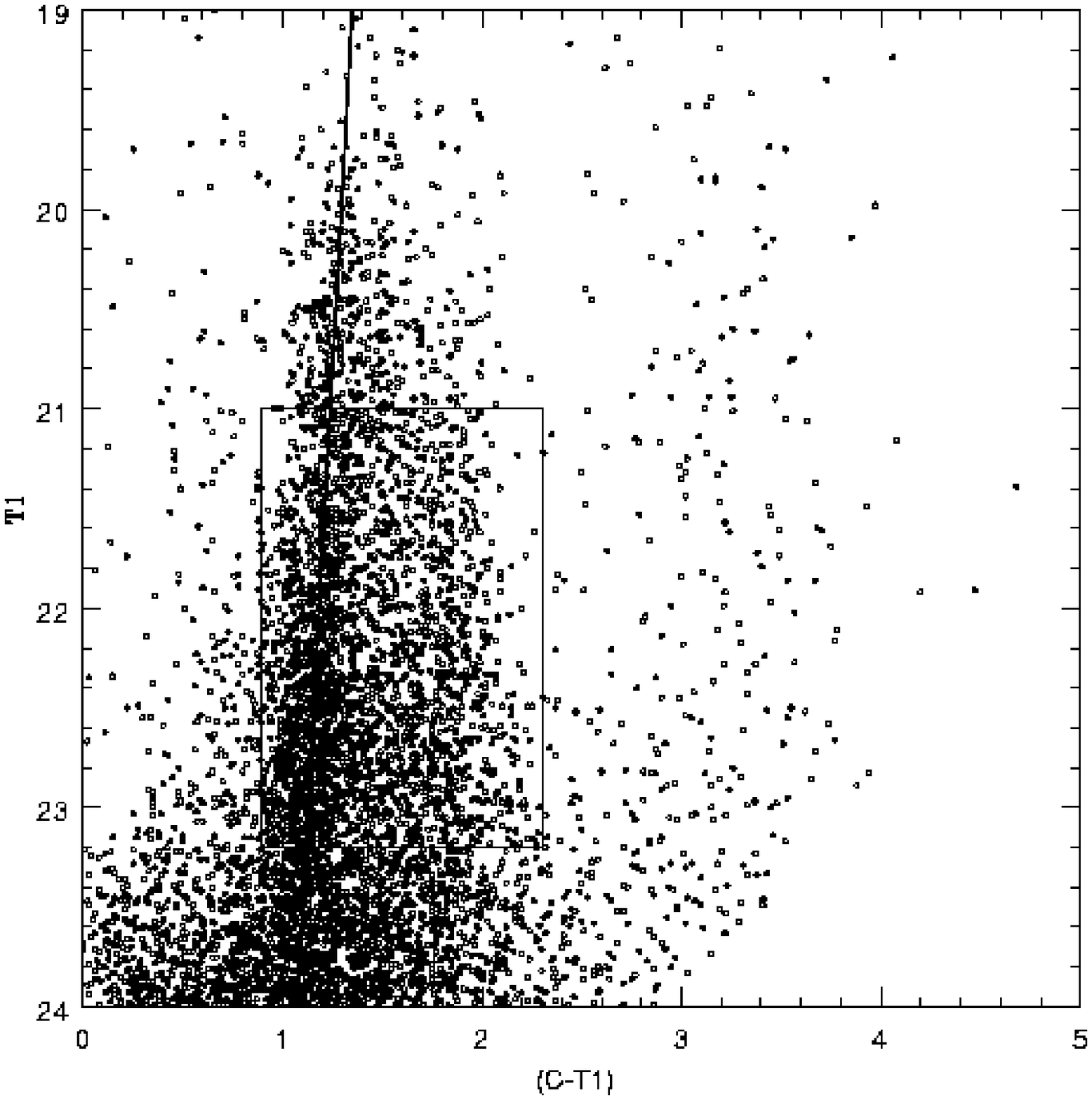}}
\caption{T$_1$ vs. (C-T$_1$) colour magnitude diagram for unresolved objects in the field of NGC 1399 (upper panel) and NGC 4486 (lower panel). The domain of the GCs candidates discussed in the text is indicated by the rectangular area. The tilted line, for NGC 4486, is defined by the modal values of the colour statistic inside 0.5 mags. intervals in T$_1$ for the blue GCs.}
\label{CMD_1399_4486}
\end{figure}

 The T$_1$ vs. (C-T$_1$) colour diagrams for non-resolved  objects 
 are displayed in Figure \ref{CMD_1399_4486}. The limiting 
 magnitude T$_1=23.2$ is indicated as a horizontal line while vertical
 lines at (C-T$_1$)=0.90 and (C-T$_1$)=2.30 define the domain of the
 globular cluster candidates.

  A distinctive feature in the lower panel of Figure \ref{CMD_1399_4486}, 
  in contrast with the upper panel of the same figure, is
  a noticeable tilt of the colours of the blue clusters associated
  with NGC 4486, which is not detectable in the case of NGC 1399. The
  tilted line Figure \ref{CMD_1399_4486} (lower panel) was obtained by 
  fitting the modal values on a 
  smoothed image of the colour magnitude diagram (adopting a round Gaussian
  kernel of 0.05 mags.) yielding:

\begin{equation}
T_1=41.50 -16.67 (C-T_1)
\label{Tilt}
\end{equation}

   This relation implies a 0.06 mags. (C-T$_1$) colour increase per 
   magnitude that is comparable to that detectable in the (g-z) colours
   of 0.045 per magnitude \citep{b77}. The possible reason for the presence
   of a ``blue'' tilt in the colour magnitude diagram is discussed in
   Section \ref{CN4486}.

\begin{figure*}
\resizebox{0.4\hsize}{!}{\includegraphics{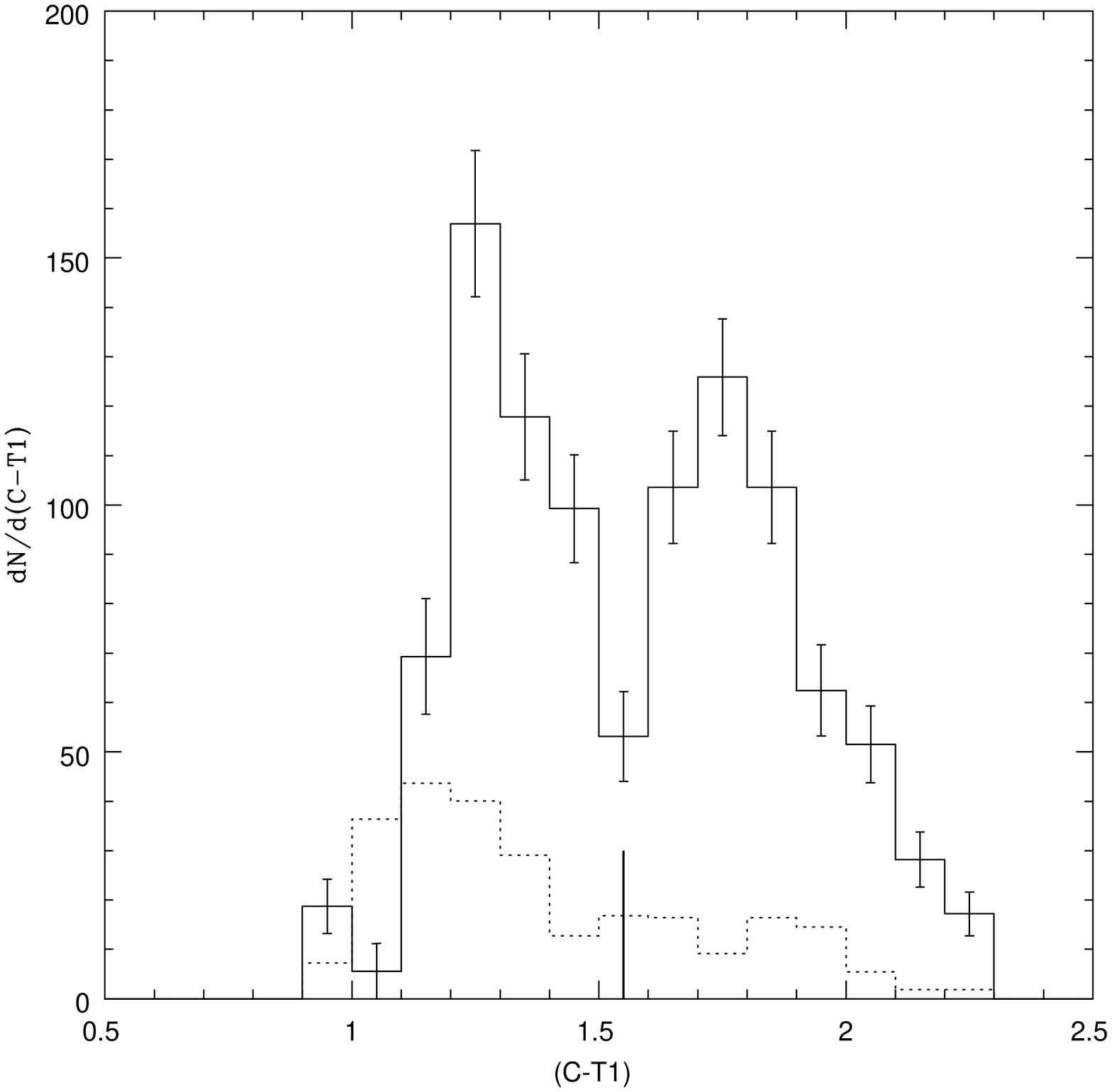}}
\resizebox{0.4\hsize}{!}{\includegraphics{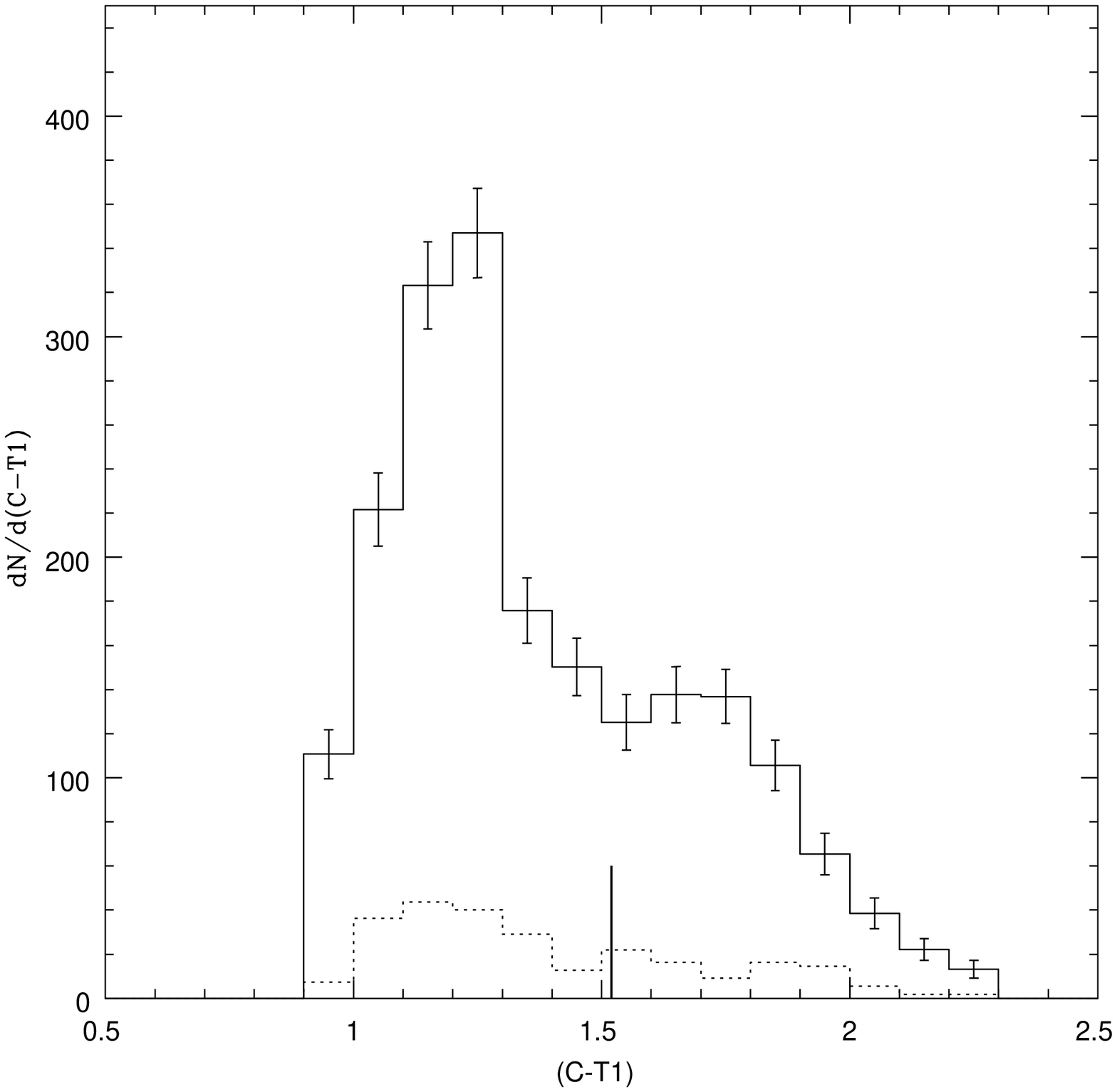}}
\caption{ (C-T$_1$) background corrected colour histogram for NGC 1399 (left panel) and NGC 4486 (right panel) globular candidates within a circular annulus defined between 120 and 360 arcsec in galactocentric radius. In both cases the areal scaled subtracted background is shown by the dotted line histograms. The combined counting statistical error bars are also shown. The histograms contain $\approx 1000$ and $\approx 1800$ cluster candidates for NGC 1399 and NGC 4486, respectively. Vertical lines idicate the position of the colour valleys.}
\label{Histos}
\end{figure*}

   The (C-T$_1$) colour histograms for GCs within a circular
   galactocentric region defined between 120 and 360 arcsec are 
   depicted in Figure \ref{Histos}. These histograms have 
   been corrected for background contamination
   by subtracting the comparison field histogram (scaled by area
   and also shown in these figures) and contain about 1000 and
   1800 candidate GCs for both galaxies, respectively.

   In order to minimise  the presence of bright objects that might
   be identified as compact dwarfs (see \citealt{b53b}) we
   adopted an upper cut off at T$_1$=21.0. We stress that \citet*{b52}
   noted that NGC 1399 cluster candidates
   brighter than this magnitude exhibit a unimodal colour distribution,
   a feature later confirmed in \citet{b14}. As a reference we point out
   that Omega Cen-like objects would appear at T$_1$ $\approx$ 21.4
   and 21.0 for NGC 1399 and NGC 4486, respectively.

   These last figures also indicate the position of the so called 
   ``colour valleys'' at (C-T$_1$)=1.55 and 1.52 for NGC 1399 and NGC 4486
   respectively. These values were determined using Gaussian smoothed
   colour histograms with a colour kernel of 0.05 mags. The same 
   procedure leads to (C-T$_1$)=1.26 and 1.21 for the blue peaks and
   (C-T$_1$)= 1.75 and 1.72 for the red peaks in both galaxies.    

\section{Description of the model}
\label{DMOD}

 In this section we describe each of the steps involved in the model
 an the main hypothesis behind it, namely:

\begin{enumerate}
\item[a)] The decomposition of the colour histograms in terms of the cluster 
 subpopulations leading to their [Z/H], [Fe/H] and (C-T$_1$) colour distributions.

\item[b)] The determination of the projected areal density distribution
 for each of the cluster subpopulations.

\item[c)] Establishing the link between each cluster and its associated
 diffuse stellar population.

\item[d)] Deriving the parameters that determine the shape of the predicted
 galaxy surface brightness profile.
\end{enumerate}

\noindent a) The decomposition of the colour histograms.

 The first step is the decomposition of the the observed colour histograms
 shown in Figure \ref{Histos} avoiding an {\it a priori} functional dependence
 (e.g., the usual Gaussian assumption). It must be emphasised that, matching the two-peaked colour histograms observed both in NGC 1399 and NGC 4486
 through the adopted colour-metallicity relation (see below),
 necessarily requires {\bf two distinct} globular cluster populations.``Seed''
 clusters were then
 randomly generated in the abundance Z domain according to a given
 statistical dependence. Trial and error shows that exponential
 behaviours {\bf $f(Z) \approx \exp [-(Z-Zi)/Zs]$} where {\bf $Zs$} is the {\bf abundance scale
 length} and {\bf $Zi$} is the minimum abundance, provides  acceptable fits to 
 the observed histograms (i.e., within the 
 Poissonian uncertainties associated with each statistical bin).
 Some more complex functions cannot be rejected
 but would imply a larger number of free parameters not justified
 in terms of those uncertainties. 

As for the minimum abundance, we adopted $Zi=0.003 Z_{\odot}$, that corresponds to $[Fe/H]=-2.65$ in the 
 metallicity scale presented below.
As discussed later, a dependence of $Zi$ with T$_1$ leads to a blue tilt that reproduces the colour-magnitude diagram of the NGC 4486 GCS.

 The decomposition procedure aims at matching the position of the
 colour peaks and colour valley while keeping a minimum value of the
 quality index of the fit, $\chi^2$ , defined as in \citet{b10b}.

 The cluster abundance $Z$ was linked to metallicity on the Zinn
 and West (1984) scale. The adoption of the [Fe/H]zw scale imply some caveats 
 (see for example, \citealt{b82} or \citealt{b76}) about the nature of this 
 index. In this work we use  the relation found by \citet*{b46b}
 for the stellar population models given by \citet*{b82b} :

\begin{equation}
[Fe/H]zw = [Z/H]-0.131
\end{equation}

  An integrated colour was then obtained for each cluster through an empirical
  (C-T$_1$)-[Fe/H] colour-metallicity calibration.

 Several approach have been made in the past
 aiming at determining the colour metallicity relation. For example, the original linear
 relation found by \citet{b26} for MW clusters was later 
 improved by \citet{b33}.
 Being an important step in the modelling process, we attempted
 a new calibration, described in Section \ref{CMC}, which
 yields a quadratic relationship between metallicity and integrated
 cluster colours.

  Before comparing the model cluster colours with the observed histograms,
  we added interstellar reddenings, ($E(B-V)=0.015$ and $0.022$ for NGC 1399
  and NGC 4486, respectively) from the \citet{b70} maps, 
  adopting $E(C-T_1)= 1.97 E(B-V)$, and Gaussian errors matching their behaviour
  as a function of cluster brightness displayed in Figure \ref{P_error}.

\begin{figure}
\resizebox{1.0\hsize}{!}{\includegraphics{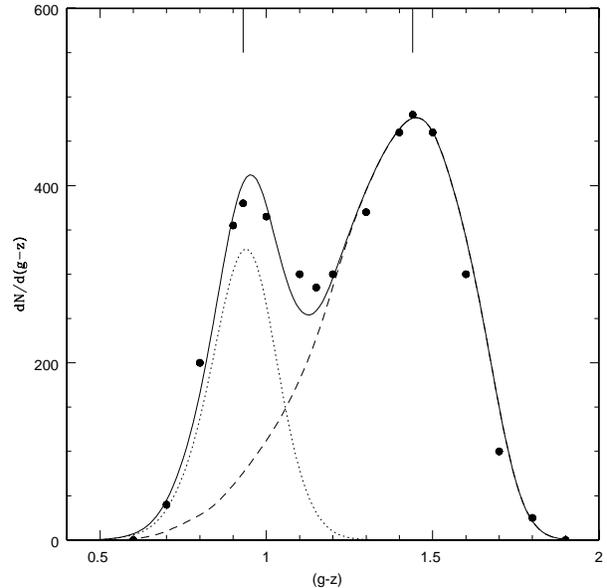}}
\caption{ Model fit (continuous line) to the kernel histogram representative of globular  clusters associated with the brightest ellipticals (black dots) in the Virgo ACS by \citet{b53}. The adopted Gaussian colour 
  kernel is 0.05 mags. The two components correspond to $Zs$ of $0.035 Z_\odot$ and $1.05 Z_\odot$ for the blue (dotted line) and red (dashed line) clusters subpopulations respectively with peaks at (g-z)=0.92 and 1.43 (vertical lines).}
\label{Model_f1}
\end{figure}

\begin{figure}
\resizebox{1.0\hsize}{!}{\includegraphics{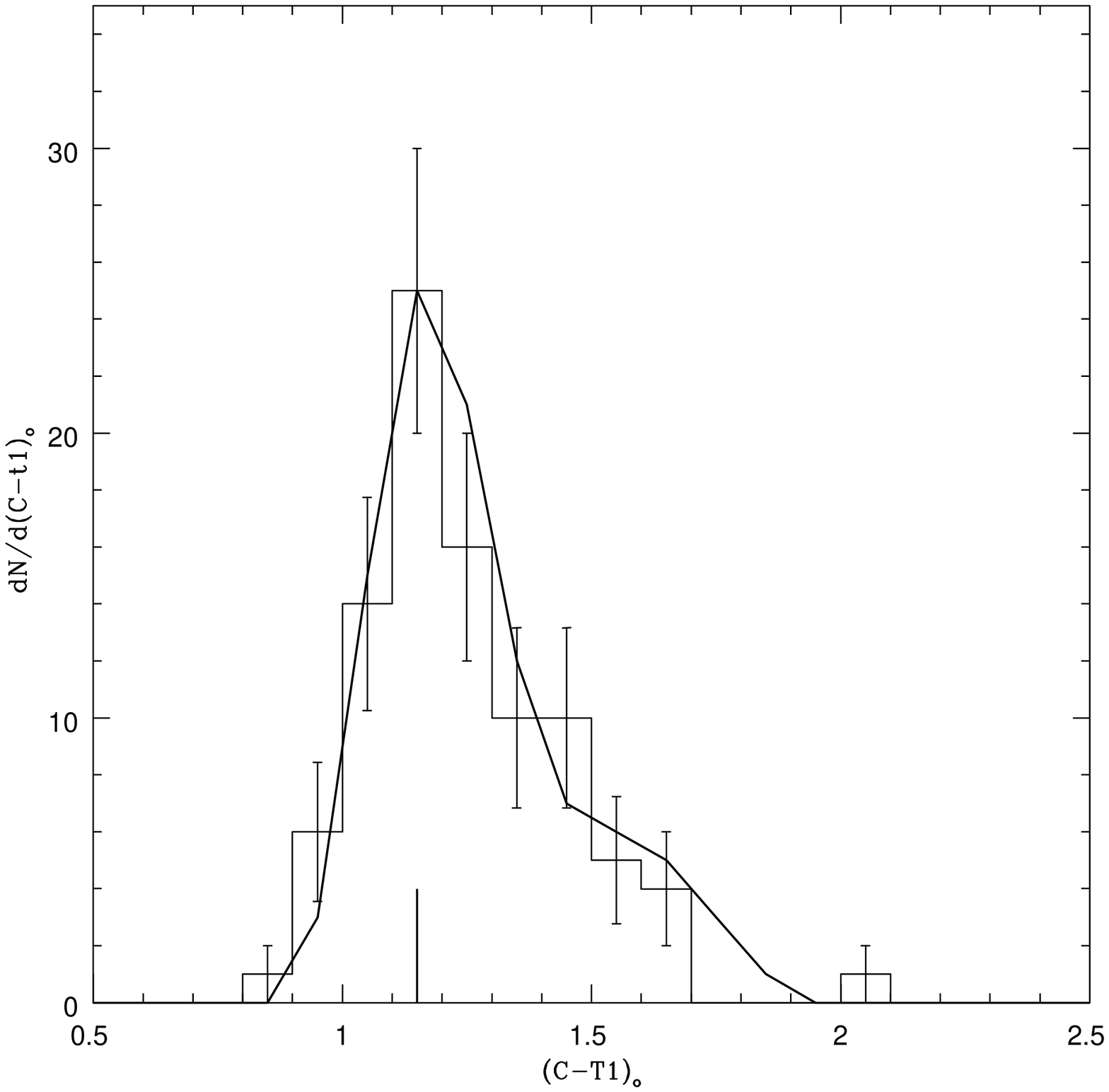}}
\resizebox{1.0\hsize}{!}{\includegraphics{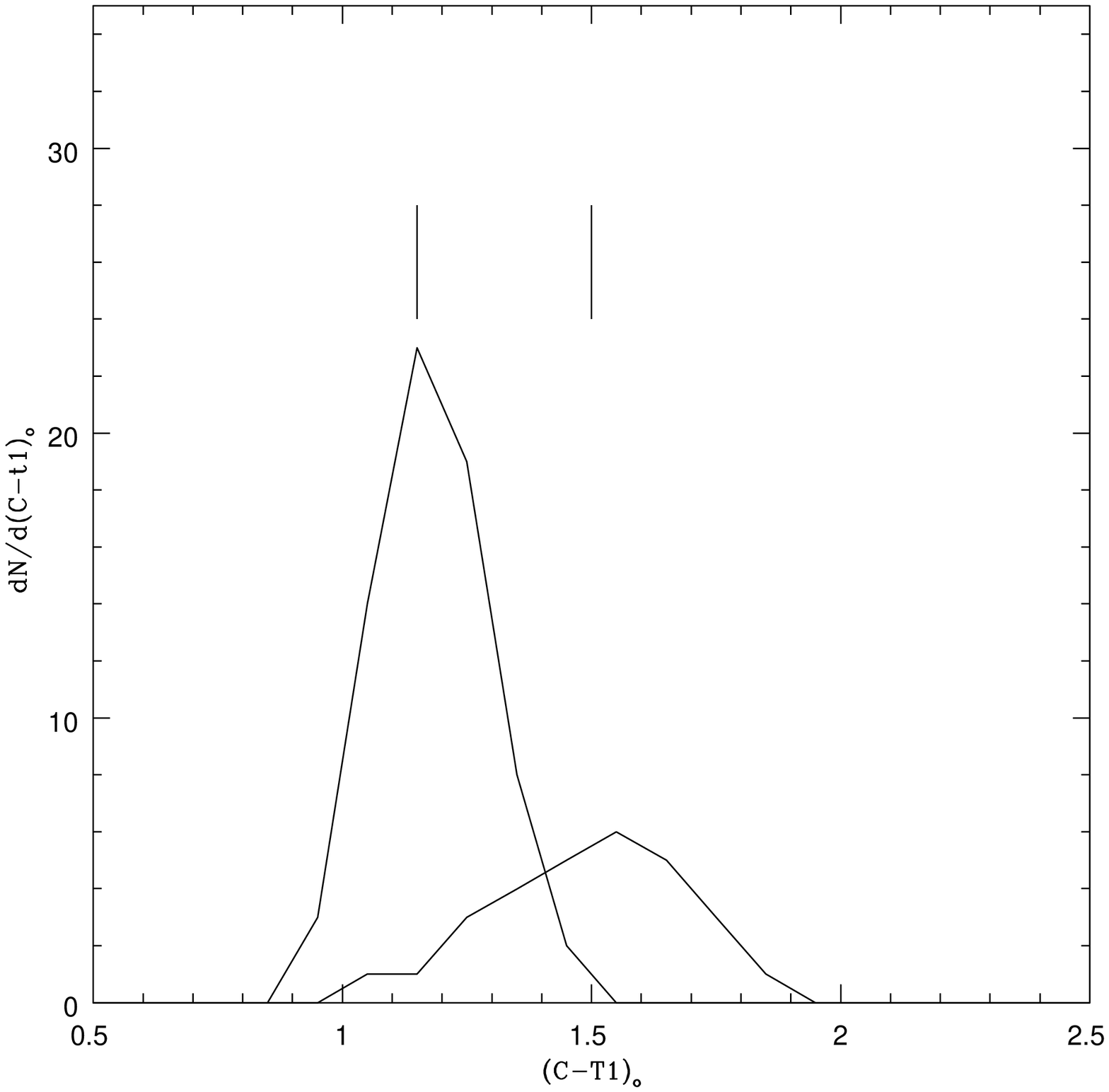}}
\caption{Upper panel: Colour histogram for 91 MW globular clusters. The bars indicate the statistical count uncertainties. The continuous line is a model fit described in text. Lower panel: Model components for MW globulars. Blue clusters are fit with an abundance scale $Zs(blue)=0.035 Z_\odot$ and red globulars with $Zs(red)=0.50 Z_\odot$. Colour peaks at $(C-T_1)_o=1.15$ and  1.51 are indicated by vertical bars.}
\label{Model_f2}
\end{figure}

  Examples of the decomposition procedure are shown in figures \ref{Model_f1} 
  and \ref{Model_f2}.
  The first diagram displays the results obtained from fitting the kernel
  colour distribution of GCs belonging to the brightest galaxies
  sample in the Virgo ACS (\citealt{b53}; figure 5). These galaxies are 
  comparable in brightness to both NGC 1399 and NGC 4486. The parameters
  of the best fit are $Zs(blue)=0.035 Z_\odot$ for the blue clusters (23\% 
  of the total population) and $Zs(red)=1.05 Z_\odot$ for the red clusters.
  In this
  case, we transformed the (C-T$_1$) colours to (g-z) by adopting: 

\begin{equation}
(g-z)= (C-T_1)-0.29
\label{CCR}
\end{equation}

  This colour transformation is consistent with the colour of the peaks
  in the ACS bright galaxies sample compared to our estimate of the 
  (C-T$_1$) peaks in Figure \ref{Histos} and also with the colours relation
  derived from the \citet{b44} models.

   Figure \ref{Model_f2} shows the best fit obtained for 91 MW globulars
   with (C-T$_1$) colours (from \citealt{b29}) or (B-I) colours
   (transformed according to $(C-T_1)_o= 1.03 (B-I)_o-0.43$) in \citet*{b61}.
   In this case, 70\% are assigned to the blue clusters
   subpopulation, with $Zs(blue)=0.035 \pm 0.01 Z_\odot$, and the remaining 30\%
   to the red one, with $Zs(red)=0.50 \pm 0.05 Z_\odot$. These parameters imply
   [Fe/H] peaks at -1.7 and -0.5, respectively. The small sample of MW
   clusters has large statistical uncertainties but the fit is consistent
   with the observed shape of the [Fe/H]  distribution (see \citealt{b4} and 
   references therein).\\

\noindent b) Projected spatial distributions.

  The model assumes that each of the GCs subpopulations has 
  its own and distinctive spatial distribution. As noted before,
  however, the adoption of a  given colour window to define each cluster 
  subpopulation, leads to ambiguous results in the case of NGC 1399.   
  This particular aspect is discussed with more detail in Section \ref{GCPADD}
  on the basis of the photometry presented in this paper. In particular,
  we stress that the variation of the slope observed for the blue
  GCs (depending on the colour window adopted as their domain)
  may arise as a partial superposition of the two cluster subpopulations.\\ 

\noindent c) The GCs-diffuse stellar population link.

  \citet{b86} introduced the {\bf T} parameter, defined as
  the total number of GCs per galaxy stellar mass unit.
  In this work, we generalise that parameter
  by assuming that the number of globular 
  clusters per associated diffuse stellar mass t is a function of total
  abundance [Z/H].  After exploring different possible functions, we adopted:
  $t=\gamma \exp(-\delta[Z/H])$ (i.e. t increases when abundance decreases),
  and then:

\begin{equation}
dN/d[Z/H]=t([Z/H]) M([Z/H])
\label{f1}
\end{equation}
  
\noindent where dN is the number of CGs associated with a stellar mass M and
 an abundance [Z/H] that belongs to a given subpopulation.        
  This assumption leads to a ``diffuse'' stellar mass per cluster with a given [Z/H]:

\begin{equation}
M^*=1/t
\label{f2}
\end{equation}

\noindent  and then to an integrated luminosity:

\begin{equation}
 L=M^*/(M/L)
\label{f3}
\end{equation}

\noindent Where (M/L) is the mass-luminosity ratio characteristic for stars 
  with the same age and metallicity of the ``seed'' globular cluster.
 
  In this work we adopted the (M/L) ratio for the B (Johnson) band given
  by \citet{b83} and an age of 12 Gy. This ratio can be approximated
  as

\begin{equation}
(M/L)_B= 3.71 + ([Z/H] + 2.0)^{2.5}
\label{f4}
\end{equation}

  This approximation differs from the Worthey's (M/L) ratios by, at most,
 $\approx 7 \%$.

  Note that we adopt [Z/H] instead of [Fe/H] as Worthey's models 
  assume solar scaled metallicities. A comparison with other models
  gives an idea about the uncertainty in this ratio. For example, models
  in \citet{b44}, for the same age and a Salpeter luminosity function,
  show an overall agreement better than $10 \%$ with Worthey's except at the
  lowest abundance where they deliver a ratio $\approx  24 \%$ larger.
  The effect of age variations on this ratio is discussed below.

  In particular, we choose the B and R bands since large scale surface
  photometry is available for both NGC 1399 and NGC 4486 
  (see Sections \ref{CN1399} and \ref{CN4486}) and no comparable data
  has been published in the C and T$_1$ bands for both galaxies.

  Although the $(M/L)_B$ ratio depends on age, we stress that most works 
  (e.g. \citealt{b37}) have not detected significant age differences
  for the cluster subpopulations in NGC 4486. In turn, \citet{b19} 
   (and also see \citealt{b55}; \citealt{b56} or \citealt{b35}) find 
  arguments to support the presence of a fraction of ``intermediate age''
  clusters in NGC 1399 and in other galaxies. However, age differences as
  large as $\pm$ 2 Gy will not
  have a strong impact on the integrated colours.\\

\noindent d) The shape and colour of the galaxy surface brightness profile.

  Each stellar mass element associated with a given ``seed'' GC (and determined
  by the adopted $\gamma$ and $\delta$ parameters)
   was split into a number of ``luminous'' particles (i.e. 100
  per cluster). These particles were statistically located on the plane of the sky by
  adopting the same spatial distribution that characterises each of the
  cluster subpopulations in order to construct bi-dimensional blue 
  image (2 arcsec per pixel) of the galaxies. A red image was also obtained by
 transforming the (C-T$_1$) colour of each luminous particle to (B-R) by means of:

\begin{equation}
(B-R)_{KC}=0.704(C-T_1)+0.269
\label{f5}
\end{equation}

\noindent  empirically obtained from MW GCs with Johnson \citep{b61}
and Washington \citep{b29} photometry.

  The synthetic B and R$_{KC}$ images were then analysed with the 
  task ELLIPSE within IRAF 
  in order to derive surface brightness profiles and colour gradients along
  the semi-major axis of the galaxies
  and, in the case of NGC 4486, the variation of ellipticity along the
  same axis.

  This treatment generalises the profile expression given in FFG05:

\begin{eqnarray}
\mu_B= (V-Mv)_o+A(B)+2.5log\big[ S_B(red) \big] \nonumber  \\
-2.5log \big[ N(red)+\frac{N(blue)}{C_B} \big]
\end{eqnarray}

\noindent where N(blue), N(red) are the areal densities of the blue and
   red clusters
   at a given galactocentric distance and $C_B= S(blue)/S(red)$.
    Introducing the definition of the
   {\bf t} parameter given before, leads to:

\begin{eqnarray}
Sn^{-1}=\frac{1.0}{\gamma} \int_{[Z/H]_l}^{[Z/H]_u}{ \frac{dn}{d[Z/H]}  \frac{1.0}{(M/L)} \exp{\{\delta [Z/H]\}} d[Z/H]} 
\label{Sn}
\end{eqnarray}
                 
\noindent where the integrals are performed on the abundance domains covered
   by each cluster family and $dn/d[Z/H]=N^{-1} dN/d[Z/H]$, being N the projected areal density of each GCs subpopulations at a given galactocentric radius, and $dn/d[Z/H]$ comes from the histogram decomposition.  
  Note that the Sn values
  do not change with galactocentric radius and are solely determined by
  the Z distribution parameters of each cluster and associated stellar
  subpopulation (which we also call ``blue'' and ``red'' in what follows). 

  Both $\gamma$ and $\delta$ were iteratively changed in order to derive a
 surface brightness profile that minimises the rms of the
  residuals when confronted with the observed profiles at
  galactocentric distances larger than 120 arcsec. This last value
  was adopted since both GCS exhibit flat spatial density cores that
  contrast with the peaked shape of the galaxy surface brightness.

  These cores can be understood as the result of 
  gravitational disruption effects that change the original population
  of GCs in the inner regions of  galaxies (see, for
  example, \citealt{b9}, and references therein) and, presumably, become less
  important for cluster orbits with larger perigalactic values.

\section{Colour-metallicity Calibration}
\label{CMC}

 We present a new  colour metallicity relation  based on 198 clusters
 that combines revised data for MW GCs and also metallicity
 data obtained for GCs in three other galaxies: NGC 3379, NGC 3923,
 and NGC 4649 (\citealt{b55}; \citealt{b56}, and Norris et al. 2007, in prep.).
 [Fe/H] values for GCs in these galaxies were derived from
 Lick indices given in the \citet{b82b} stellar population
 models. Theses works were selected as they were homogeneous both in data
 handling and in the derivation of the Lick indices. 

 In the case of MW clusters, we first looked for a transformation of colours
 in the Johnson system to (C-T$_1$). A large photometric sample, that includes (B-I) colours,
 is available
 in \citet{b61}. In turn, (C-T$_1$) colours were obtained 
 from \citet{b29}. Intrinsic colours for these globulars 
 were then obtained by using colour excesses determined by \citet{b60},
 when available, or the \citet{b61} values. As a 
 result we obtain: 

\begin{equation}
(C-T_1)_o = 1.03(\pm 0.02) (B-I) - 0.43 (\pm 0.03)
\label{W_J}
\end{equation}

 In turn, the extragalactic GCs were observed in the (g-i) Sloan
 colour and transformed to (C-T$_1$) through two different ways.
 On one side using the  (g-i) to (B-I) relation derived from model 
 integrated colours given by \citet{b44} and then to (C-T$_1$)
 through our own transformation, leading to:

\begin{equation}
(C-T_1)_o = 1.43(\pm 0.03) (g-i) + 0.01(\pm 0.02)
\label{W_S1}
\end{equation}

 Alternatively, \citet{b65} have calibrated the Sloan indices
 in terms of Johnson's colour indices that can be transformed to
 (C-T$_1$)$_o$ (see, for example, \citealt{b20}) yielding:

\begin{equation}
(C-T_1)_o = 1.39(\pm 0.03) (g-i) + 0.01(\pm 0.03)
\label{W_S2}
\end{equation}

 As these transformations are very comparable, within errors, we adopted
 an average of both in order to obtain the extragalactic GCs colours on the (C-T$_1$)
 scale.

 The intrinsic colours for the extragalactic clusters, were derived by
 subtracting the interstellar reddening excess indicated by the \citet{b70} maps
 and assuming $E(C-T_1)=1.97 E(B-V)$. 

\begin{figure}
\resizebox{1.0\hsize}{!}{\includegraphics{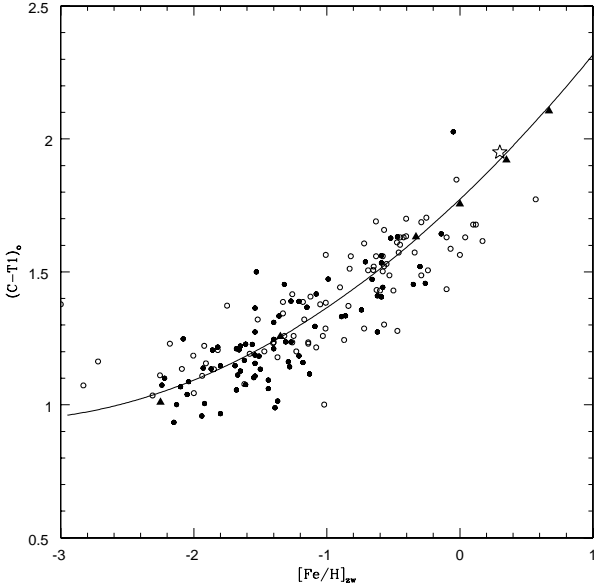}}
\caption{ (C-T$_1$) colour versus metallicity ([Fe/H] on the Zinn and West scale) relation derived from Milky Way clusters (filled dots) and globulars in NGC 3379, NGC 3923 and NGC  4649 (open dots). The star represents the nucleus of NGC 1399, triangles are models from \citet{b44} (see text). The continuous line is a quadratic fit adopted as the mean calibration.}
\label{W_Fe_fig}
\end{figure}

 The adopted colour-metallicity relation is 
 displayed in Figure \ref{W_Fe_fig}, where a quadratic fit gives a good
 representation of the data:
                  
\begin{equation}
(C-T_1)_o = 0.94 + 0.068 ([Fe/H]_{zw} +3.5)^2
\label{W_Fe}
\end{equation}

  The non-linear nature of this relation has been noted by other authors (e.g. 
  \citealt{b33}; \citealt*{b41b}) and a linear fit to the data displayed in this
  last figure leaves significant colour residuals both at the high and low metallicity
  regimes.

\begin{figure}
\resizebox{1.0\hsize}{!}{\includegraphics{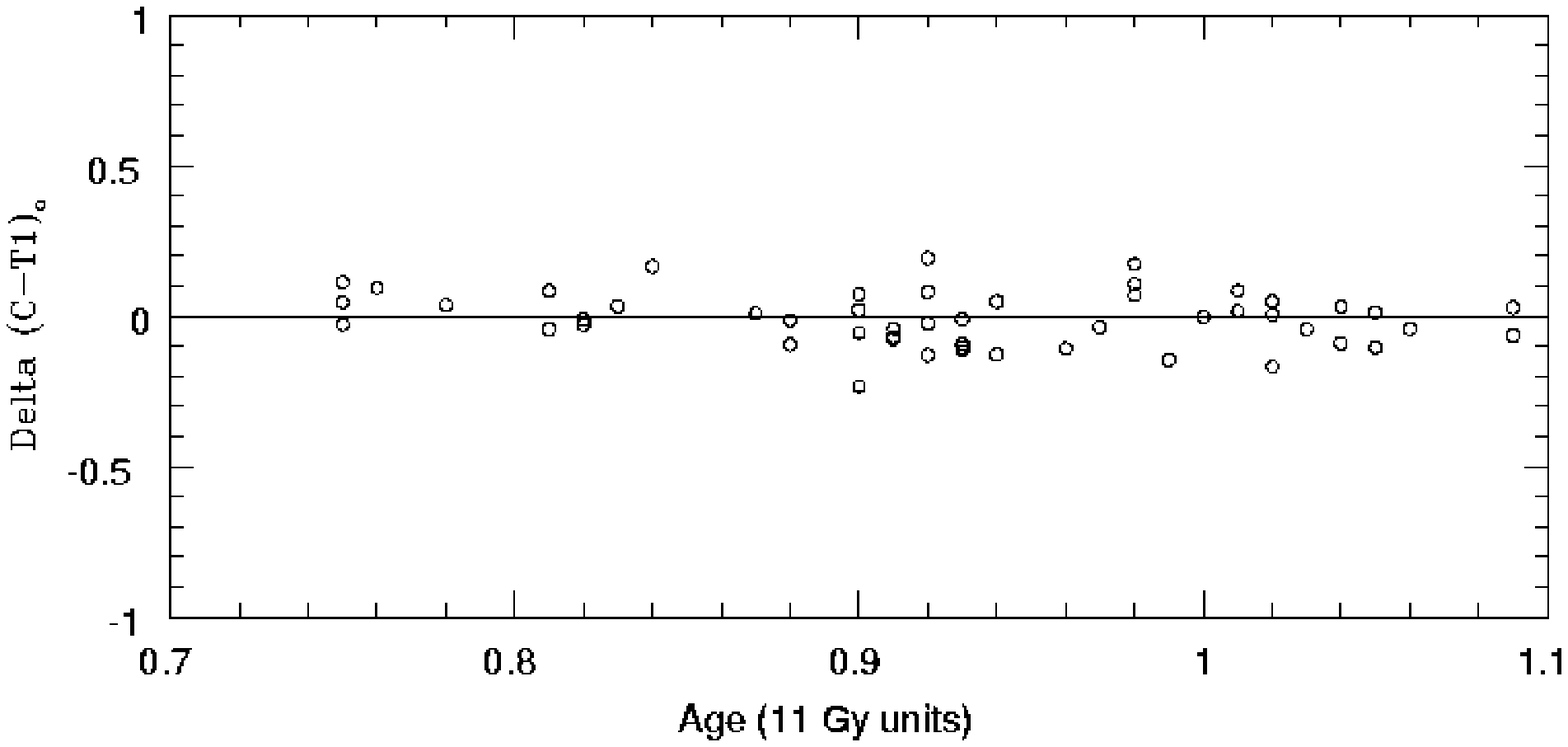}}
\resizebox{1.0\hsize}{!}{\includegraphics{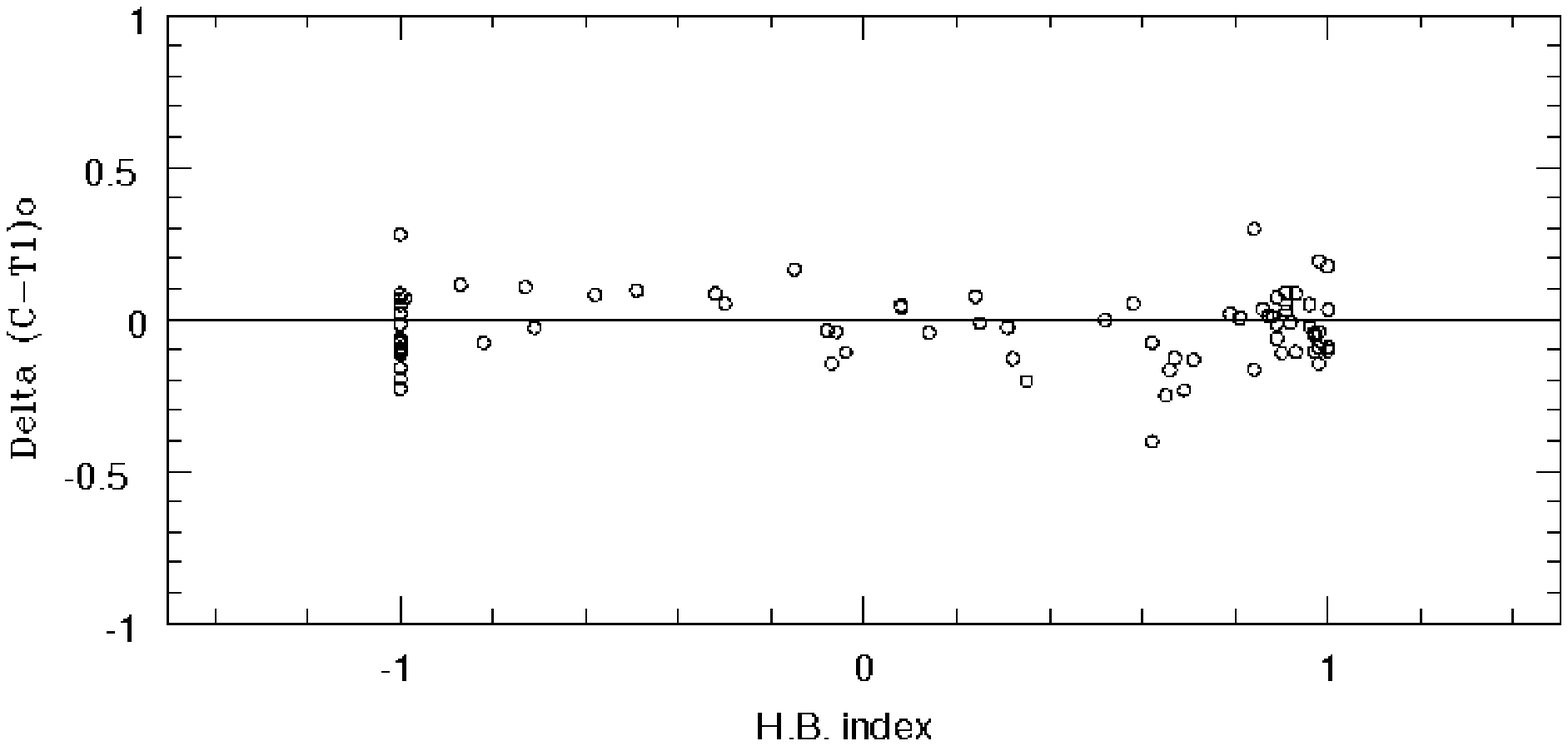}}
\caption{Upper panel: $(C-T_1)_o$ colour residuals from the mean colour-metallicity calibration for MW clusters included in Figure \ref{W_Fe_fig} as a function of normalised ages available in \citet{b13}. Lower panel: (C-T$_1$)o colour residuals for MW clusters included in Figure \ref{W_Fe_fig} and horizontal branch morphologies (HB index) available in \citet{b43}. No significant trends are detectable}
\label{W_Fe_res}
\end{figure}

  An analysis of colour residuals as a function of age (for globular with ages
  available in \citealt{b13}) or horizontal branch morphology,
  through the HB-index given by  \citealp{b43}, reveals no trends
  with these quantities as displayed in Figure \ref{W_Fe_res}. These results are in agreement with 
  a similar analysis presented by \citet{b73}, who discuss those effects for a number of different 
  colour indices.

We note that the shape of the empirical calibration is similar, to
within $\pm$ 0.015 mags. in (C-T$_1$), with the colours of the 12 Gy model,
with a Salpeter luminosity function and blue horizontal branch,
 given in \citet{b44}. This agreement is reached after subtracting a zero
point difference of 0.065  mags. to their (B-I) model colours and then
transforming to (C-T$_1$) through the relation given above. Note that these models
(shown as triangles in Figure \ref{W_Fe_fig}) are given as a function
 of total abundance [Z/H].

Figure \ref{W_Fe_fig} also includes, as a reference, the NGC 1399 nucleus according to the
 photometry by \citet{b52b} and the metallicity ($[Fe/H]=0.4$) obtained by \citet{b54}.

\section{The globular clusters projected areal density distributions}
\label{GCPADD}

  As shown in FFG05 (figure 5) the slope of the areal density of the
  bluest GCs (i.e., bluer than the blue peak at 
  (C-T$_1$)=1.26) in NGC 1399 is significantly shallower than that 
  corresponding to the ``whole'' blue population (i.e., all clusters
  bluer than the colour valley at (C-T$_1$)=1.55) in the {\bf inner} region
  of the galaxy. At larger galactocentric radii, these slopes become identical within the uncertainties.
  The significance of that result is analysed in this section on the 
  basis of the new data set presented in this work.

  First, we focus  on the areal density distributions of clusters in the
  inner regions of both galaxies. The size of this region was defined
  aiming at: a) including a large number of cluster candidates; b)
  keeping the overall completeness level of the sample at $\approx$ 95\% 
  ; c) minimising the fraction of contaminating non-resolved
  field interlopers (19\% and 11\% for NGC 1399 and NGC 4486, respectively). 

  These requirements are met within a circular annulus with inner
  and outer radii of 120 and 420 arcsec. Within 120 arcsec the searching
  routines are 
  affected by the galaxy halo brightness while, further out in 
  galactocentric radius,  the background level increases and
  the effective areal coverage of our images decreases.

  We also set a magnitude range (T$_1$=21.0 to 23.2) for two reasons.
  On one side,  in order to avoid the
  eventual presence of very bright objects
  whose nature might be connected with Omega Cen-like objects or compact dwarf
  galaxies (e.g. see \citealt{b53b}). On the other, the GCs colour distribution
  becomes "unimodal" in NGC 1399 (Ostrov et al. 1998) making difficult 
  a separation between blue and red GCs.

  Given the relatively small angular scale of this analysis we adopt
  $r^{1/4}$  laws in order to obtain least squares fits to the
  logarithmic surface densities within concentric circular annuli
  (one arcmin wide):

\begin{equation}
log (den) = a r^{1/4}+ b
\end{equation}

\begin{figure}
\resizebox{1.0\hsize}{!}{\includegraphics{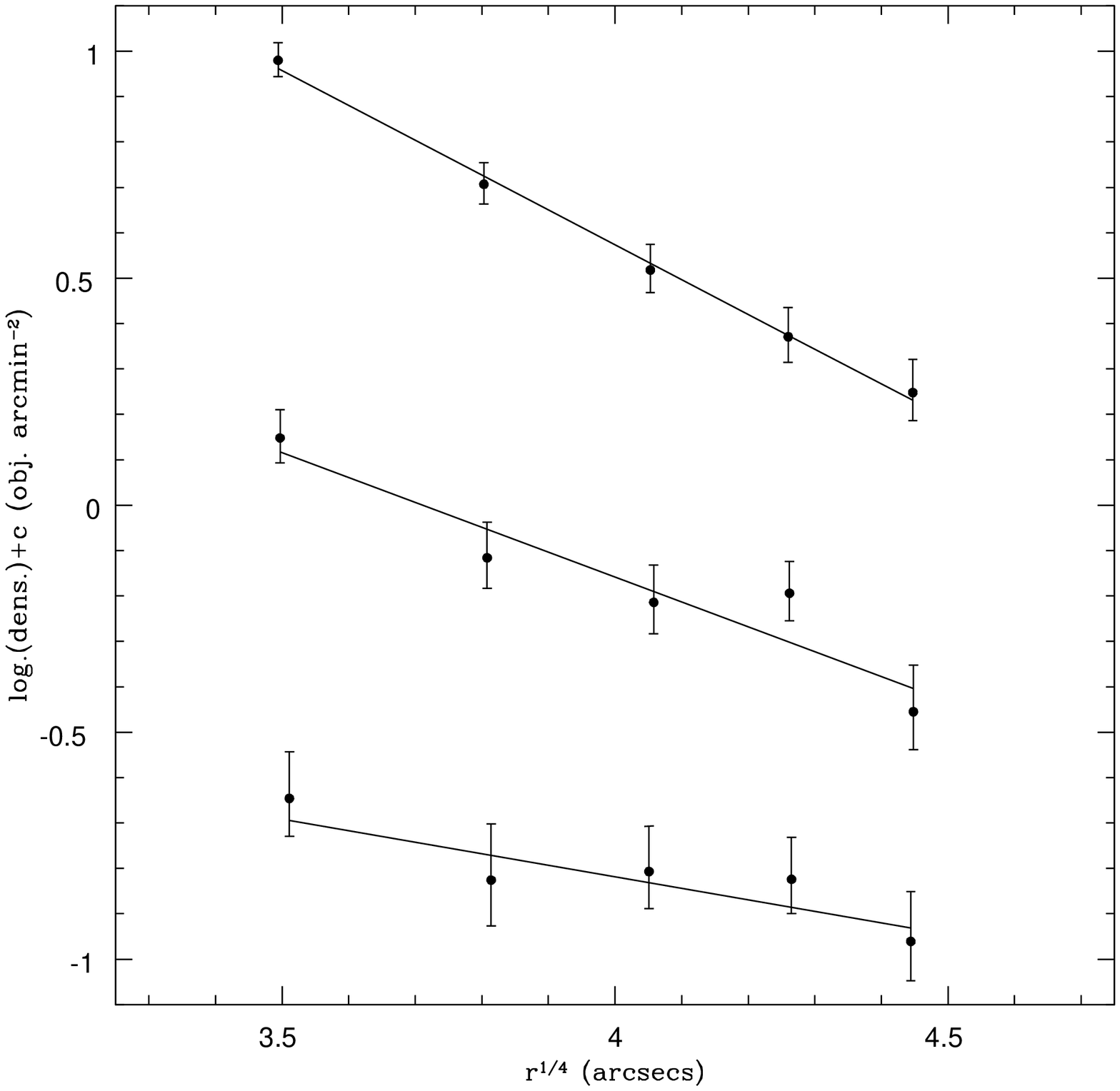}}
\resizebox{1.0\hsize}{!}{\includegraphics{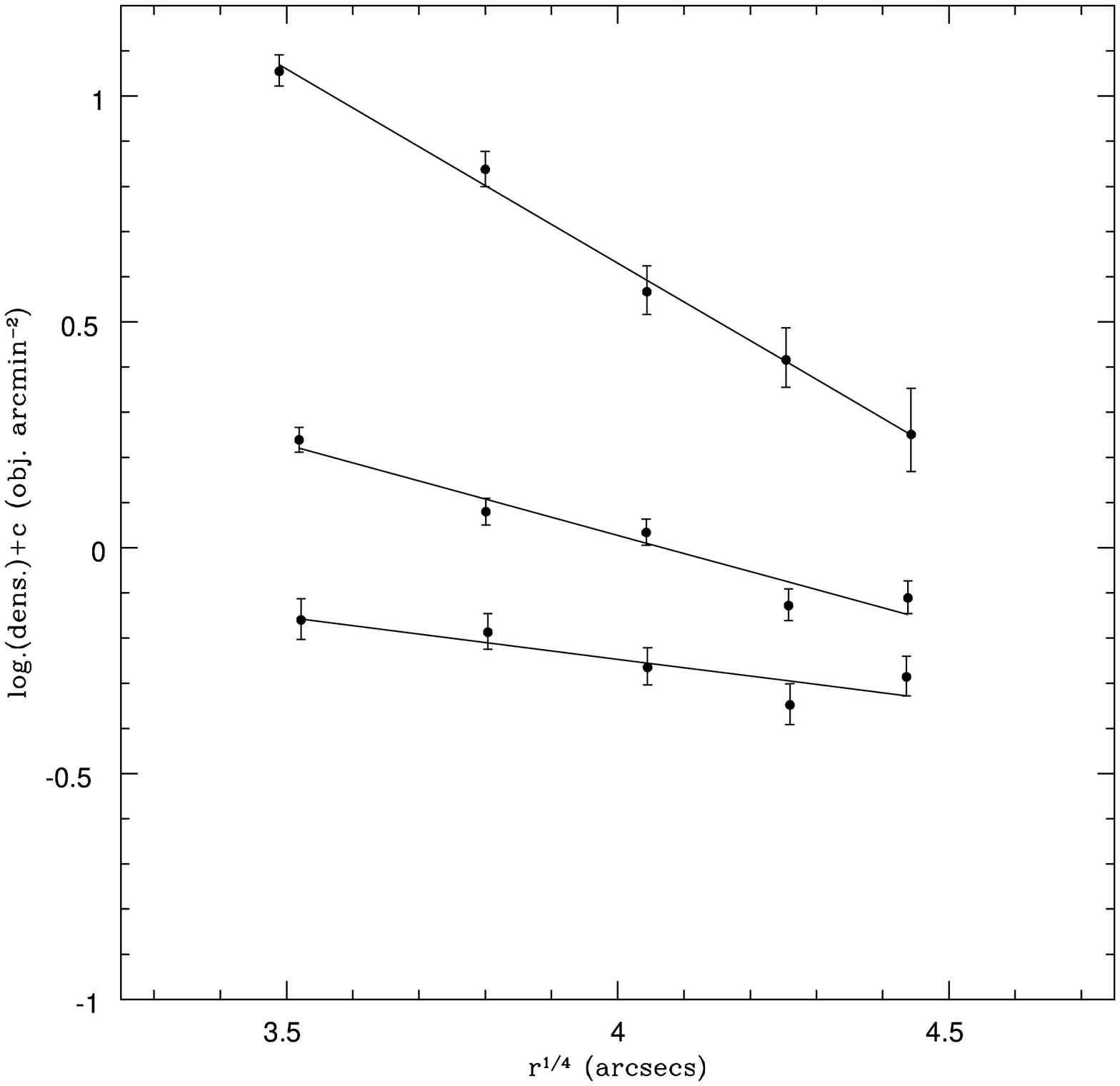}}
\caption{Upper panel: Projected areal density for clusters in NGC 1399. The upper line belongs to clusters redder than (C-T$_1$)=1.55 (``red'' globulars). The intermediate line corresponds to clusters with colours between 0.9 and 1.55 (``blue'' globulars according to the most usually adopted definition). The lowest line belongs to clusters bluer than the blue peak at (C-T$_1$)=1.26 (or ``genuine'' blue clusters according to the text). Lower panel: Projected areal density for clusters in NGC 4486. The upper line belongs to clusters redder than (C-T$_1$)=1.52 (``red'' globulars). The intermediate line corresponds to clusters with colours between 0.9 and 1.52 (``blue'' globulars according to the most usually adopted definition). The lowest line belongs to clusters bluer than the blue peak at (C-T$_1$)=1.21 (or ``genuine'' blue clusters according to the text).}
\label{Pend_1399_4486}
\end{figure}

  The resulting slopes and their associated uncertainties are listed in
  Table \ref{fits_values} and depicted in Figures \ref{Pend_1399_4486}.
  The upper two fits, in each panel, belong to the red
  and blue globulars defined in terms of the colour valleys at (C-T$_1$)=1.55
  and 1.52 for NGC 1399 and NGC 4486, respectively. These fits, that correspond
  to the regions with the highest  GCs areal densities,  are later
  overlapped (Figures 11, 12 , 17 and 18) with profiles that extend to larger
  galactocentric radii.

\begin{table}
\caption{$r^{1/4}$ law fits to logarithmic areal densities for globular clusters (T$_1$=21.0 to 23.0)}
\label{fits_values}
\begin{tabular}{|c|c|c|c|}
\hline
\multicolumn{1}{c}{colour range} &
\multicolumn{1}{c}{a (slope)} &
\multicolumn{1} {c} {b (zero point)}  &
\multicolumn{1} {l} {rms} \\
\hline
\multicolumn{1}{c}{NGC 1399} &
\multicolumn{1}{c}{ } &
\multicolumn{1}{c}{ }  &
\multicolumn{1} {c} { }  \\
1.55 - 2.30  &   -0.77  $\pm$ 0.03  &   3.64 $\pm$ 0.11  &  0.02 \\
0.90 - 1.55  &   -0.43  $\pm$ 0.10  &   2.28 $\pm$ 0.41  &  0.08 \\
0.90 - 1.26  &   -0.25  $\pm$ 0.11  &   1.20 $\pm$ 0.43  &  0.08 \\
\multicolumn{4} {c} { }  \\
\multicolumn{1}{c}{NGC 4486} &
\multicolumn{1}{c}{ } &
\multicolumn{1}{c}{ }  &
\multicolumn{1} {c} { }  \\
1.52 - 2.30 &   -0.91  $\pm$   0.08   &  4.26  $\pm$   0.34  &  0.06 \\
0.90 - 1.52 &   -0.46  $\pm$   0.04   &  2.87  $\pm$   0.20  &  0.04 \\
0.90 - 1.21 &   -0.22  $\pm$   0.03   &  1.64  $\pm$   0.12  &  0.02  \\
\hline
\end{tabular}
\end{table}

  In turn, the lower fits belong to GCs bluer than the respective blue peaks
  (at C-T$_1$=1.26 and 1.21). Both galaxies show very similar behaviours in 
  in the sense that the bluest globulars exhibit significantly shallower
  density slopes as found in FFG05 for the case of NGC 1399 but using
  Washington photometry from \citet{b14}.

  A further analysis, adopting different colour windows, shows no
  meaningful differences in the slopes of the  clusters redder
  than the colour valley, and we consider that they ``genuinely'' belong to
  a single population.

  The intermediate slope value observed for the  whole blue GCs
  sample (compared to the bluest GCs) tentatively  suggests 
  that an overlap between the blue and red globular subpopulations may occur
  in the colour range defined between the blue peak and the colour valley.
  This overlap would increase the density slope of the {\bf so far} called 
  blue clusters as result of the presence of the blue tail of the red
  subpopulation within their formal domain (i.e., objects bluer than the
  colour valley).
 
  That effect should decrease with increasing galactocentric radius, as 
  the presence of the red clusters becomes less prominent due to the 
  steeper density profile of these clusters. This tentative picture is
  discussed in the following sections.

\section{The case of NGC 1399}
\label{CN1399}

1) Colour histogram decomposition.

\begin{figure}
\resizebox{1.0\hsize}{!}{\includegraphics{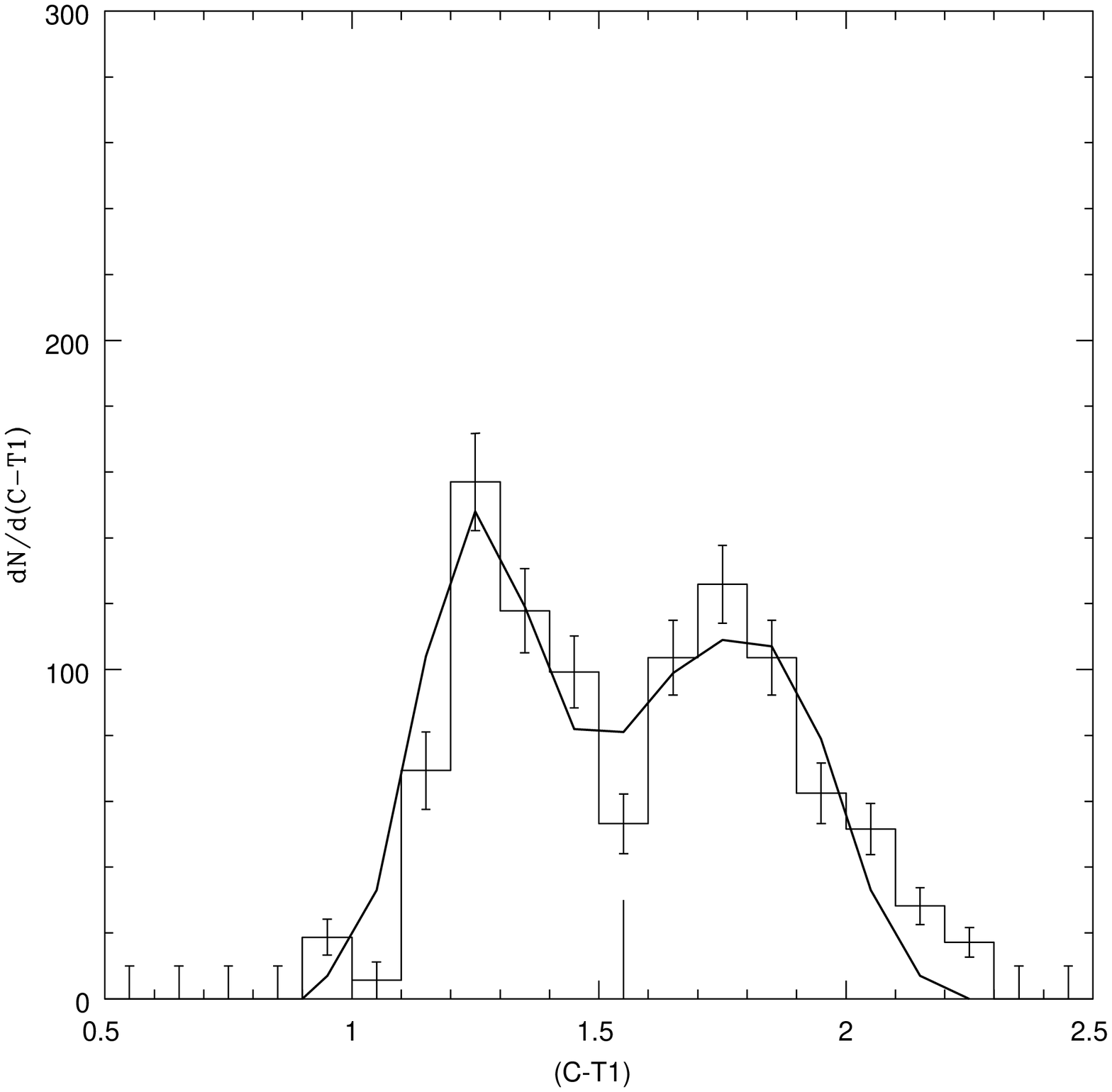}}
\resizebox{1.0\hsize}{!}{\includegraphics{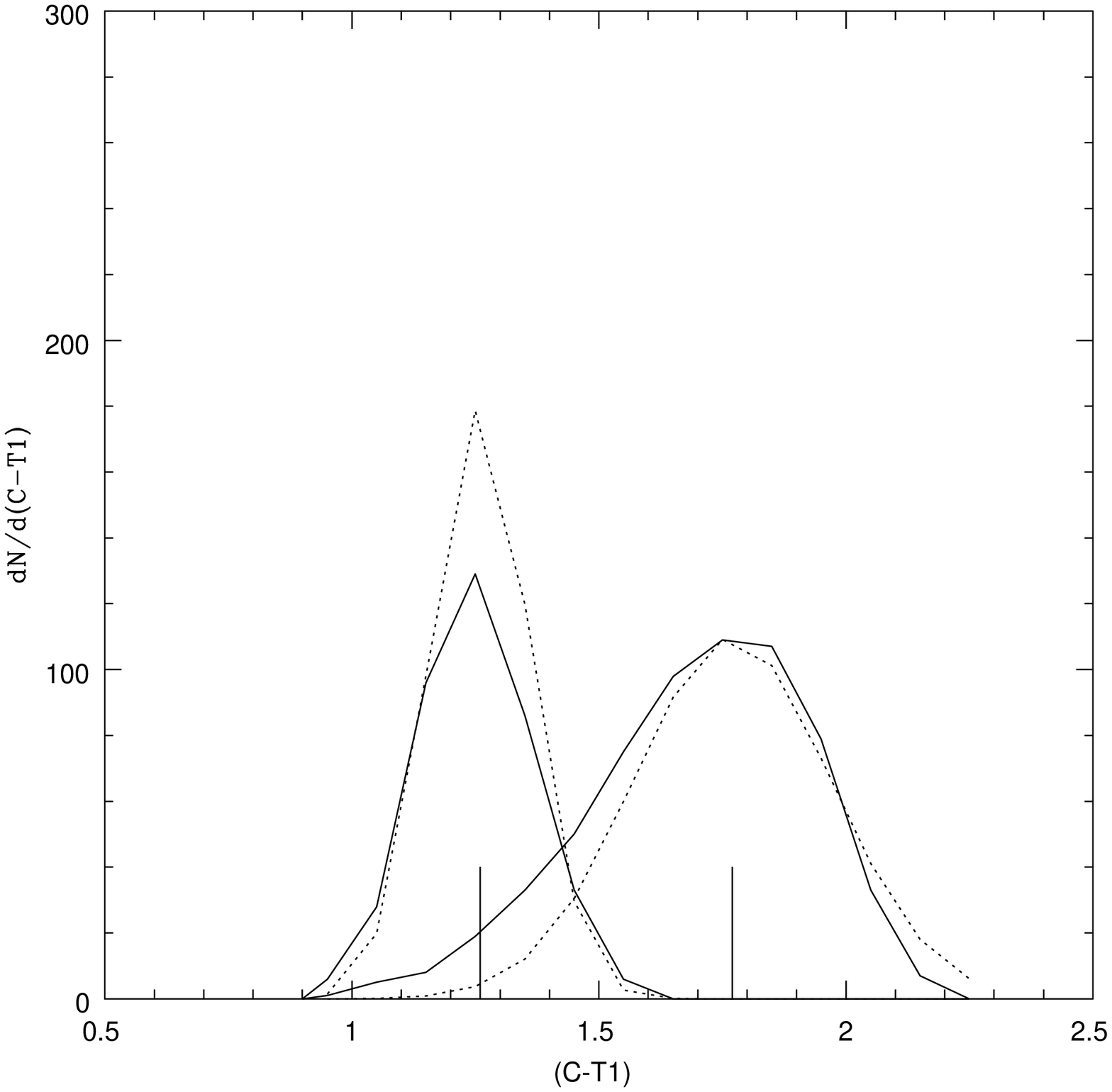}}
\caption{Globular populations fit to the background corrected colour histogram for clusters in NGC 1399 is depicted in the upper panel. The vertical line at (C-T$_1$)=1.55 is the colour ``valley'' usually adopted as the boundary between both subpopulations. The lower panel shows each of the globular populations derived from exponential distributions in the abundance domain Z. Vertical lines indicate the colours peaks of each clusters populations. Dotted lines in the lower panel show the components of the best Gaussian fit.}
\label{Model_1399}
\end{figure}

  The background corrected GCs colour histogram 
  is compared in Figure \ref{Model_1399} with a synthetic one derived 
  through the 
  modelling described in Section \ref{DMOD}. This histogram includes $\approx 
  1000$ GCs candidates with T$_1$=21.0 to 23.2 and (C-T$_1$)=0.90 to 2.30.

  The decomposition process yields 620 clusters to the red subpopulation
  with an abundance scale factor $Zs(red)= 1.45 \pm 0.1 Z_\odot$. The remaining 380 
  globulars are identified as belonging to the blue population with
  an abundance scale $Zs(blue)=0.045 \pm 0.01 Z_\odot$. Figure \ref{Model_1399}
 (lower panel), with comparison purposes, also displays the Gaussian components that give the best
 fit to the observed histograms (blue clusters: $\overline{(C-T_1)}=1.26$,
 $\sigma_b=0.12$; red clusters: $\overline{(C-T_1)}=1.77$, $\sigma_r=0.20$).
 These fits decrease the  number of red clusters and increase
 the number of blue ones  suggesting a smaller degree of colour
 overlapping between both GCs subpopulations in comparison with the results from 
 the model. 

 As discussed below, the eventual inclusion of a blue tilt comparable
 with that adopted for the NGC 4486 blue GCs does not have a detectable 
 effect on the shape of the colour histogram.

   Figure \ref{Model_1399} suggests that a single abundance scale 
   parameter for the
   red GCs population falls somewhat short in  the 
   extreme red end of the colour histogram. About 5\% of that
   population appears definitely redder than the model prediction.
   A tentative explanation might suggest some degree of field
   contamination in that colour range or a possible effect connected
   with a variation of the $[\alpha/Fe]$ ratio with age \citep{b38}.
 
   Figure \ref{Model_1399} also shows that the model colour distribution 
   of the red GCs exhibit a ``blue'' tail  (i.e., clusters bluer than the
   colour valley at (C-T$_1$)=1.55), representing about 31\% of 
   the total number of red clusters.

   These objects appear as ``contaminating'' the formal domain of the
   genuine blue globulars and will have an impact on the density slopes
   derived for this last population if only the colour valley is
   adopted as a discriminating criteria between both GCs subpopulations.

   Alternatively, the model blue GCs barely reach the
   colour valley, suggesting that this population will not affect
   the estimate of the areal density slope of the red clusters if
   only clusters redder than the colour valley are included in the sampling. \\

\noindent 2) The areal density distribution of the blue and red globulars.

   Due to the features discussed in the previous item, we only take
   clusters bluer than (C-T$_1$)=1.25 (the blue peak) as {\bf tracers} of the
   surface density of the {\bf genuine} blue GCs. Figure \ref{Model_1399} shows 
   that there would still be a small degree of contamination by the
   bluest clusters of the red population (about 5\% of the
   total sample within that colour range).

\begin{figure}
\resizebox{1.0\hsize}{!}{\includegraphics{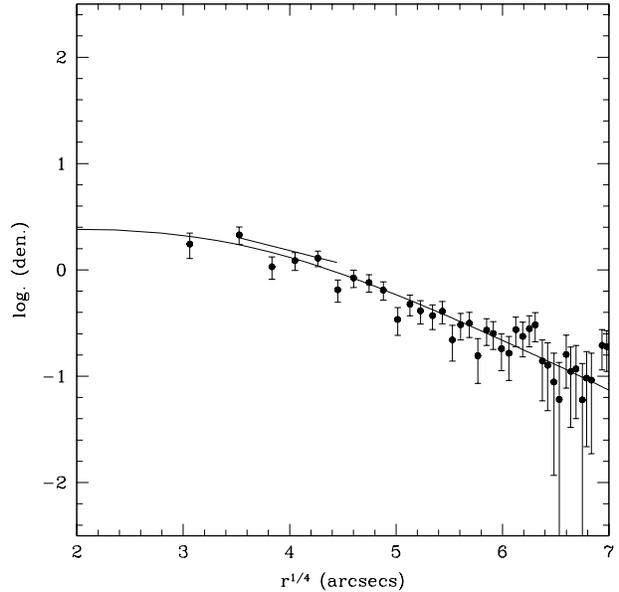}}
\caption{Large scale areal density distribution for globular cluster bluer than the blue peak at (C-T$_1$)=1.26. This colour domain is practically uncontaminated by  the blue tail of the red globulars population and considered as the real distribution of the ``genuine'' blue globulars. The filled black dots come from data in \citet{b2}. The continuous line is projected areal density described in text. The short straight line is the fit for the bluest globulars in our photometry as depicted in Figure \ref{Pend_1399_4486}}.
\label{Dens_1399_B}
\end{figure}

   The density run with galactocentric radius of these clusters is depicted in Figure \ref{Dens_1399_B}. 
   In this
   case we only include objects with T$_1$=21.0 to 23.2 taken from the
   photometric work by \citet{b2}, that reaches a galactocentric radius of
   40 arcmin. The short straight
   line represents the density fit discussed in Section \ref{GCPADD} while the
   continuous line comes from projecting on the sky a volumetric
   density profile: 

\begin{equation}
\rho(a)=C (1.0+(a/rs))^{-3}
\label{Vol_1399}
\end{equation}
                    
\noindent where $a$ is measured along the galaxy major axis,  a scale length $rs=375$ arcsec and a  spatial cut-off at a galactocentric radius of 450 kpc.

\begin{figure}
\resizebox{1.0\hsize}{!}{\includegraphics{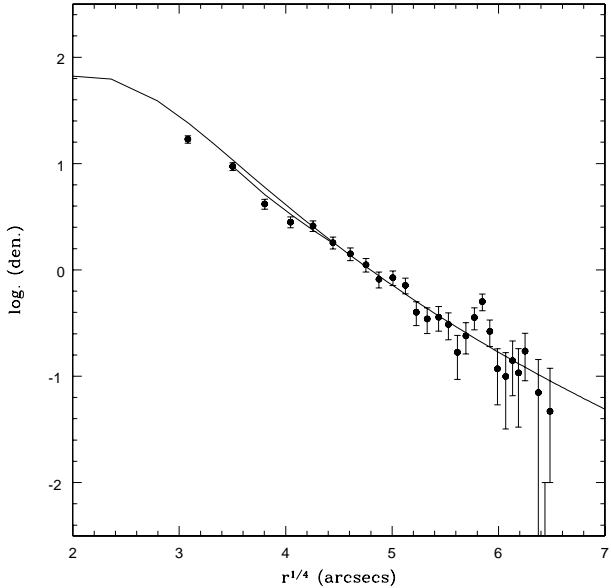}}
\caption{The same as figure 8 but for clusters redder than (C-T$_1$)=1.55. The continuous line is a Hubble profile with a 
core radius r$_c$=60 arcsec. The short straight line is the fit for the red globulars in our photometry as depicted in Figure \ref{Pend_1399_4486}.}
\label{Dens_1399_R}
\end{figure}

   The large scale density distribution for the red GCs was then derived 
   using only clusters redder than (C-T$_1$)=1.55 as tracers of that
   population and is shown in Figure \ref{Dens_1399_R}. The short straight line is the
   fit discussed in Section \ref{GCPADD} while the continuous line is a Hubble
   profile with a core radius r$_c$=60 arcsec.

   In order to estimate the density distribution of the {\bf total} number of
   clusters for each population, the density of the tracer GCs
   should be increased by factors that take into a account the total
   colour range covered by these populations (adding the blue clusters
   redder than the blue peak and the red clusters bluer than the 
   colour valley, respectively), as indicated by the colour histogram modelling, 
   and also the sampled fraction of globulars within their respective integrated
   luminosity functions.

   \citet{b28} have derived the luminosity functions of 
   both blue and red GCs populations in NGC 1399 on the basis of HST
   WFPC2 observations. Assuming fully Gaussian luminosity functions, 
   and transforming  (B-I) colours to (C-T$_1$), their results lead to turn
   overs at T$_1$=23.40 and T$_1$=23.45 with dispersions of 1.24 and 1.16 mags
   for the blue and red populations respectively.

   The combined colour and luminosity completeness factors are then, 4.28 for the 
   blue globulars and 3.29 for the red GCs.\\

\noindent 3) The surface brightness profile.

   Surface brightness photometry for NGC 1399 in the B band up to
   a galactocentric radius of 775 arsecs was presented in FFG05 and 
   compared with other profiles available in the literature (e.g.
   \citealt{b71}; \citealt*{b8}).

   The predicted blue profile was obtained  through the procedure 
   described in Section \ref{DMOD} and adopting a distance
   modulus (V-Mv)$_o$=31.4,
   corresponding to 19 Mpc (see FFG05 and references therein), and an
   interstellar colour excess E(B-R)=0.011 (transformed from \citealt{b70}).

   Azimuthal counts do not show a detectable flattening of the NGC 1399
   GCS and therefore we adopted the average flattening of the galaxy,
   q=0.86, as representative for both cluster populations.

\begin{figure}
\resizebox{1.0\hsize}{!}{\includegraphics{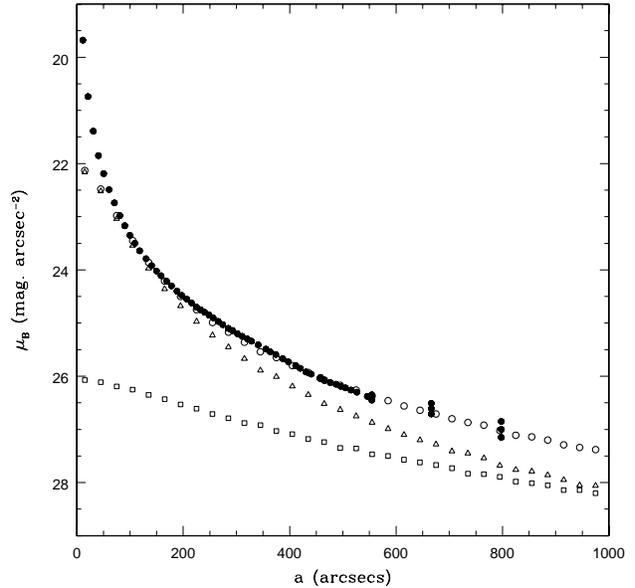}}
\caption{Observed B surface brightness profile for NGC 1399 (FFG05; filled dots) confronted with the model fit (open circles). This last profile was obtained from a bi-dimensional model image processed with the ELLIPSE task within IRAF. Squares and triangles represent the luminosity associated with the ``blue'' and ``red'' stellar populations. Note that the model fails inside 100 arcsec in galactocentric radius where the globular distributions shows a flat core.}
\label{Prof_1399}
\end{figure}

   The model surface brightness profile delivered by ELLIPSE is compared
   with the FFG05 observations in Figure \ref{Prof_1399} and corresponds 
   to $\gamma=0.82 (\pm 0.05) \times 10^{-8}$ and $\delta$=1.1 $\pm$ 0.1.
   The overall rms of the fit is $\pm$ 0.035 mags.

\section{The case of NGC 4486}
\label{CN4486}

  1) Colour histogram decomposition.

\begin{figure}
\resizebox{1.0\hsize}{!}{\includegraphics{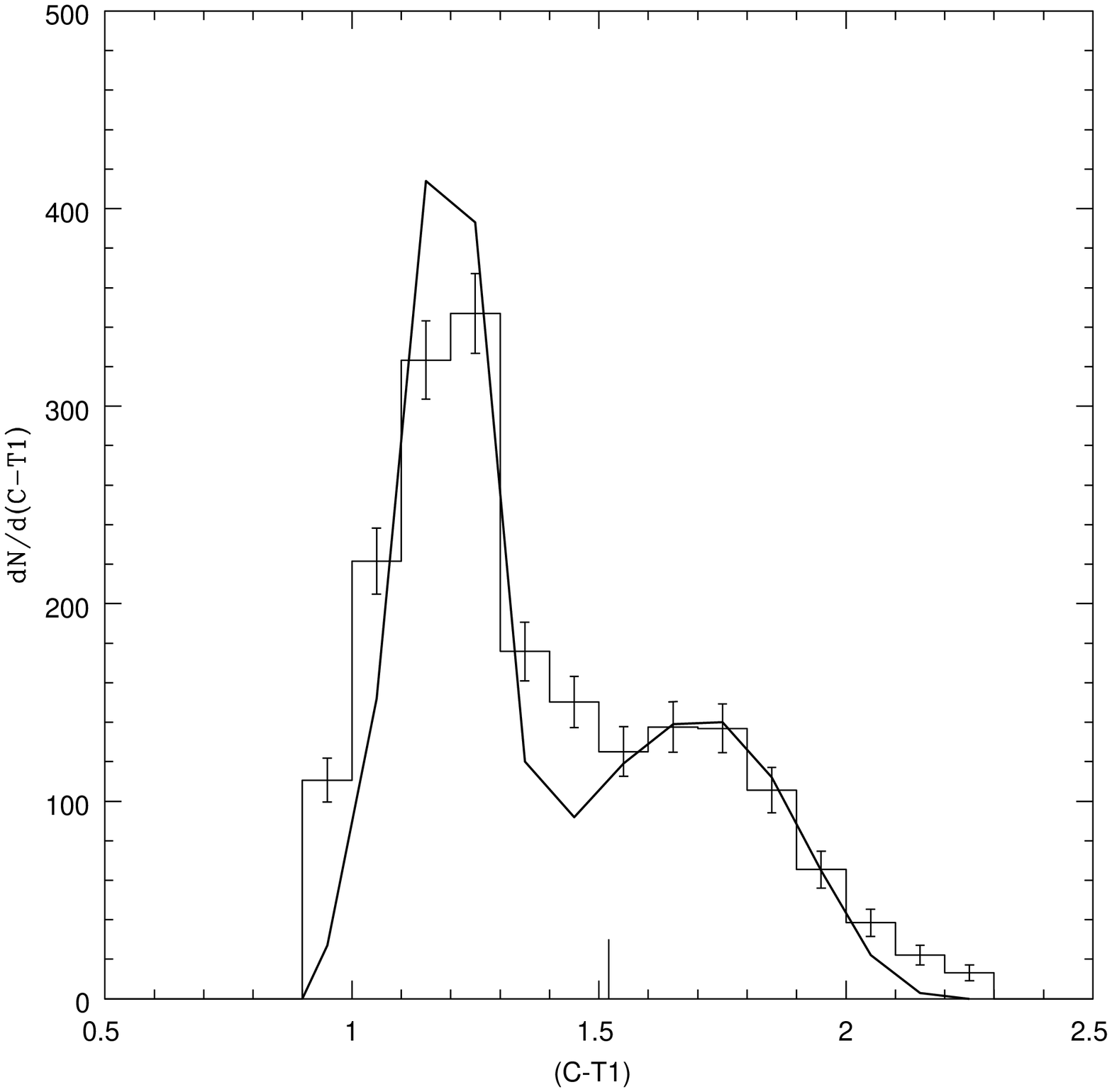}}
\resizebox{1.0\hsize}{!}{\includegraphics{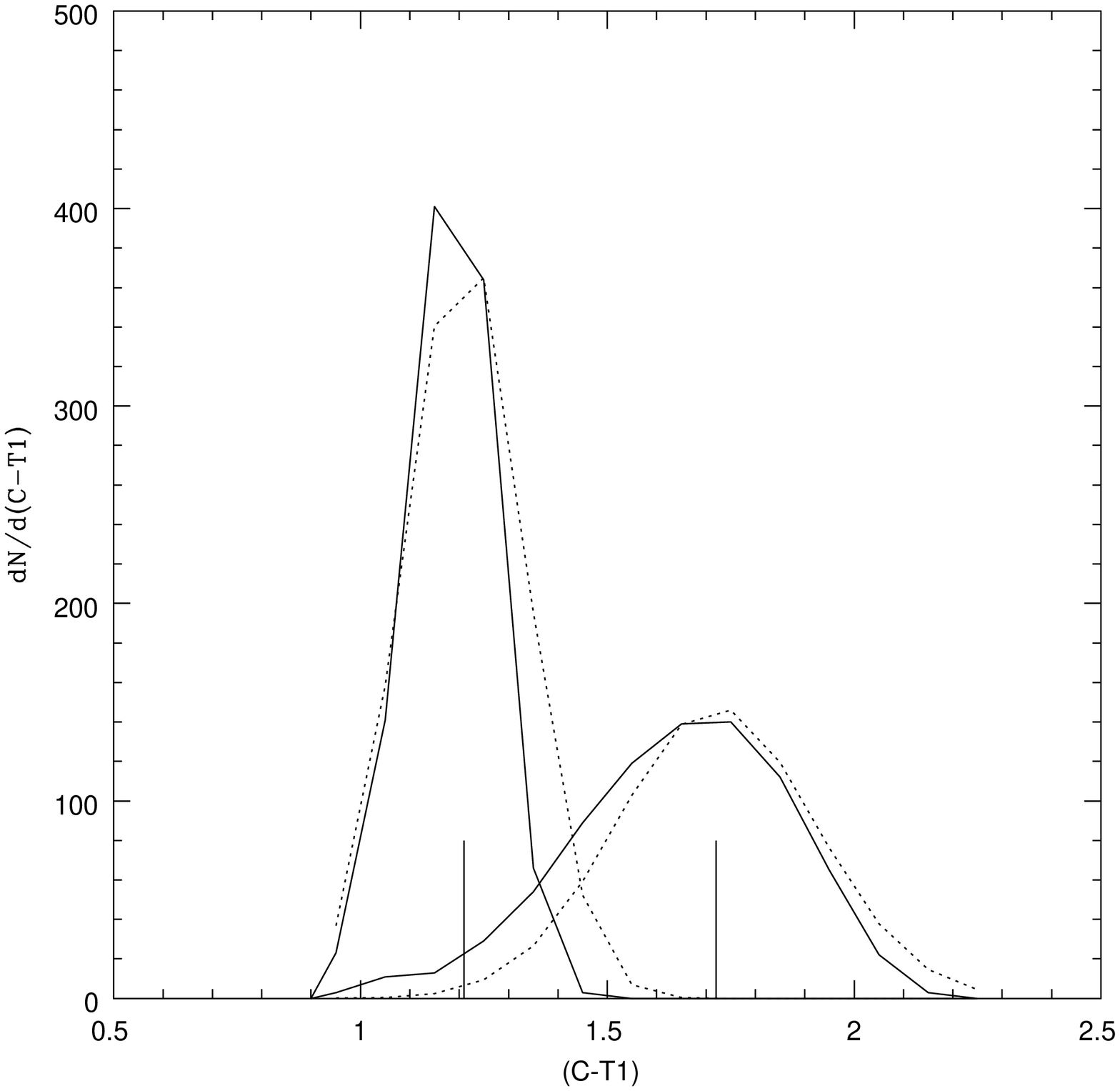}}
\caption{A globular population fit to the background corrected colour histogram for clusters in NGC 4486 (see Figure \ref{Histos}) is  depicted in the upper panel. The vertical line at (C-T$_1$)=1.52 is the colour ``valley'' usually adopted as the boundary between both populations. The lower panel shows each of the globular populations derived from exponential distributions in the abundance domain Z. Vertical lines indicate the colour peaks of each cluster populations. Dotted lines in the lower panel show the components of the best gaussian fit.}
\label{Model_4486}
\end{figure}

     The GCs background corrected colour histogram
     is compared with the model fit in Figure \ref{Model_4486}. In this 
     case, 800 clusters were assigned to the red population with an abundance
     scale $Zs(red)=0.90 \pm 0.1 Z_\odot$ and 1000 clusters to the blue
     population with $Zs(blue)=0.012 \pm 0.005$ practically unde Z$_\odot$ (see below). Figure 14
     also shows the Gaussian components
    (blue clusters: $\overline{(C-T_1)}=1.21$, $\sigma =1.12$;
    red clusters: $\overline{(C-T_1)}=1.72$, $\sigma=0.20$). Here we
     also adopt an initial abundance $Zi=0.003 Z_\odot$  for 
     the red clusters. However,
     as shown in Section \ref{CMDCD}, the blue GCs display an evident 
     tilt  that we associate with a change in abundance that correlates
     with the cluster brightness and, hence, mass (see also figure 3
     in  \citealt{b5}).  In this case, we find that a change
     in initial abundance as a function of brightness:

\begin{figure*}
\resizebox{0.4\hsize}{!}{\includegraphics{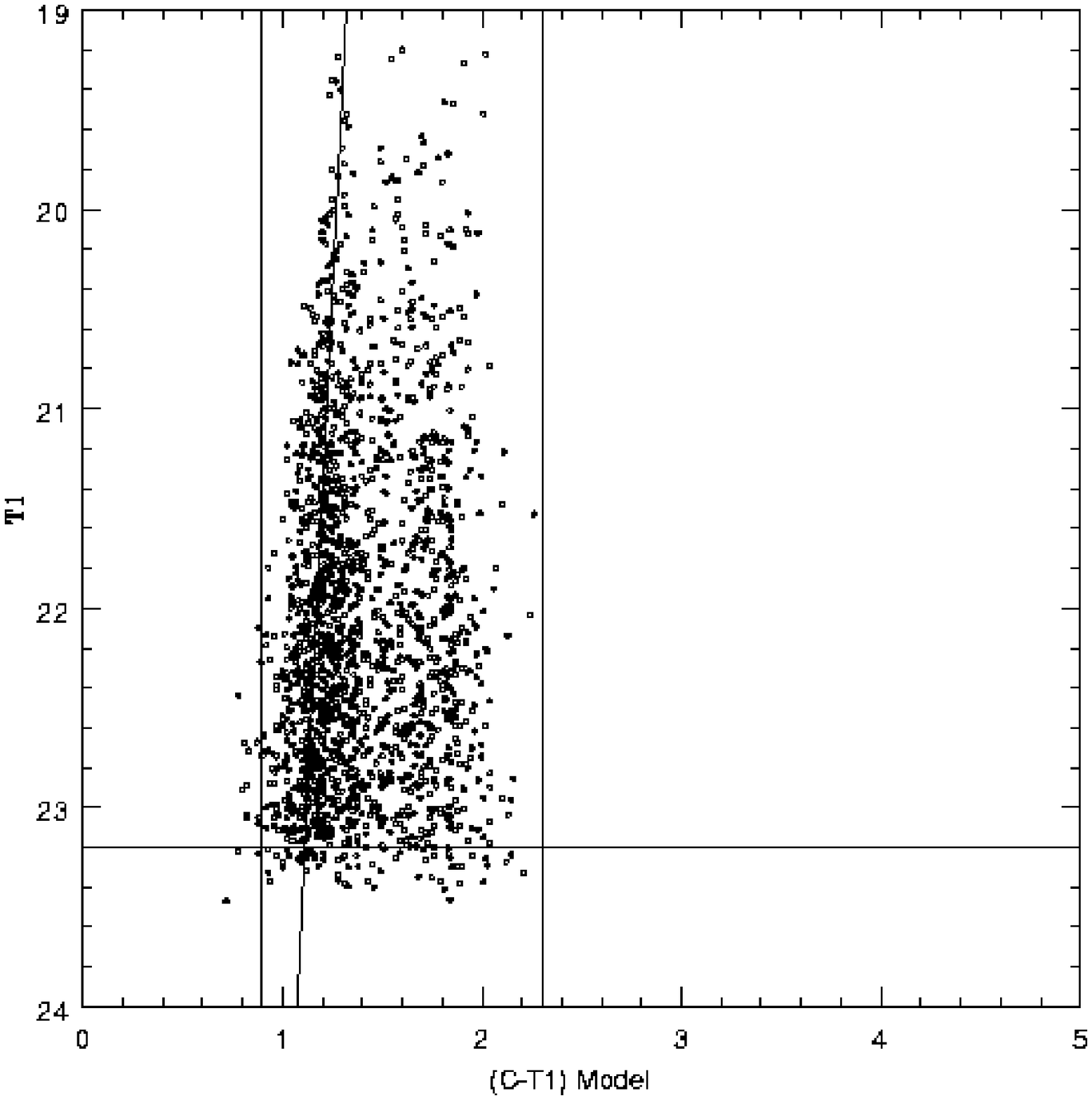}}
\resizebox{0.4\hsize}{!}{\includegraphics{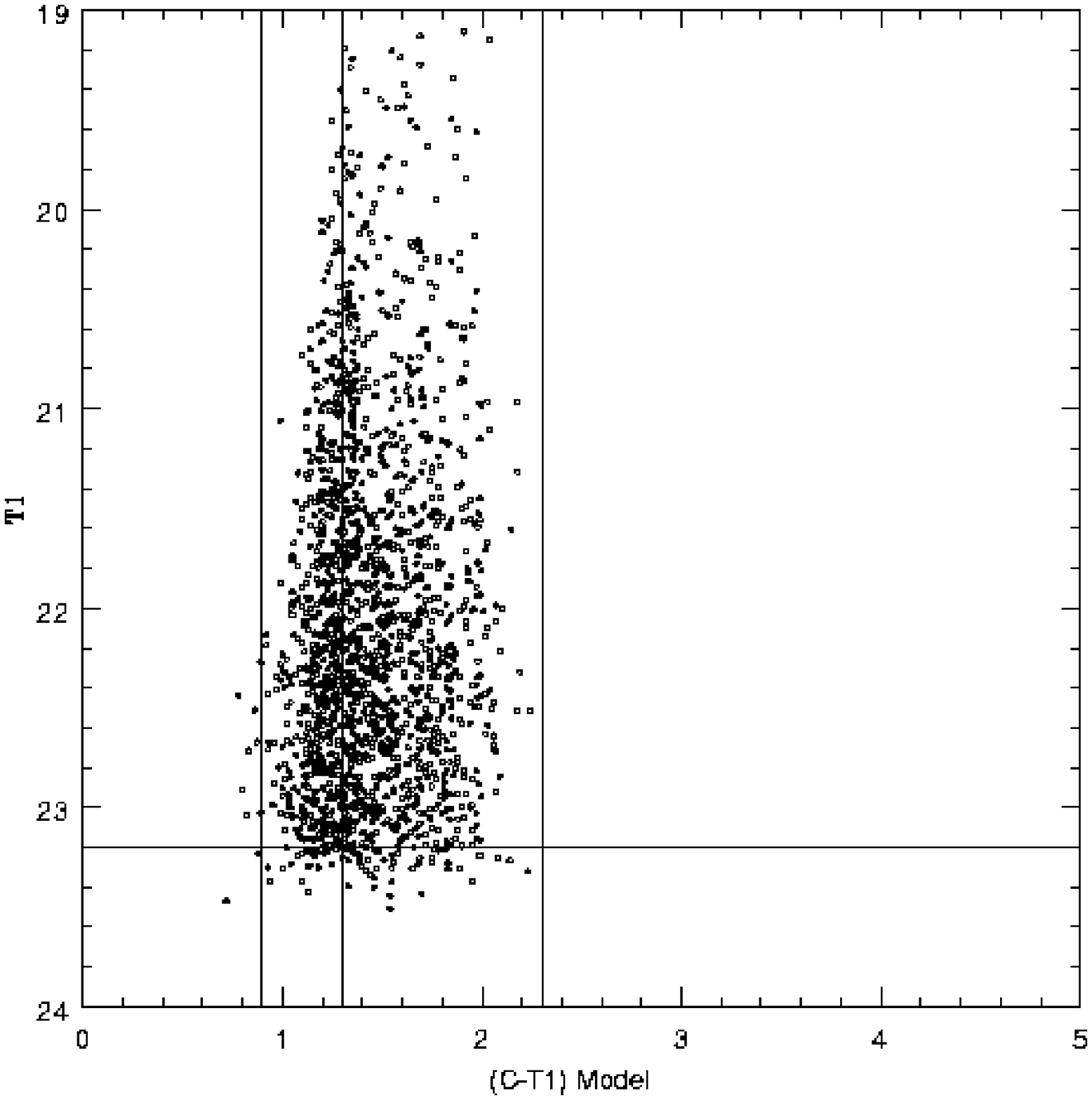}}
\caption{ Model colour diagrams with $Zs(blue)=0.012 Z_\odot$ (left panel) and $Zs(blue)=0.05 Z_\odot$. Both models include a blue tilt similar to that discussed in the text. However, the larger Zs(blue) adopted in the right panel, makes more difficult the detection of the tilt.
}
\label{Tilt_4486_1399}
\end{figure*}

\begin{equation}
\Delta Z = 0.01 (23.2-T_1)~~~~(Z_\odot ~~units)
\label{Tilt_4486}
\end{equation}

\noindent reproduces the appearance of the blue GCs colour-magnitude diagram. The
 mean Z for blue GCs with T$_1$ from 21.0 to 21.25 mags. is $0.0371 Z_{\odot}$ while for
 clusters with T$_1$ from 22.95 to 23.2 is $0.017 Z_{\odot}$. These values are
 consistent with a mass/metallicity scaling relation (where M is the clusters mass):

\begin{equation}
Z \approx M^{0.44}
\label{Z_M_4486}
\end{equation}

\noindent somewhat smaller than $Z \approx M^{0.55}$ but comparable to $Z \approx M^{0.48}$
 suggested respectively by \citet{b34} and \citet{b77}. Figure 3 also suggests that
 the blue GCs tilt, as noted by the last authors, is in fact detectable over the whole
 magnitude range brighter than T$_1$=23.2.

     As mentioned in Section \ref{CMDCD}, a tilt is not detected in the
     case of the NGC 1399 GCs. In this galaxy, blue GCs exhibit 
     a considerably larger Zs(blue) than in NGC 4486 and we suggest that,
     this larger abundance spread
     makes more difficult the detection of an eventual tilt. 
     As an example, Figure \ref{Tilt_4486_1399} displays the colour magnitude diagram
     for the adopted model in the 
     case of NGC 4486, showing the blue tilt (left panel).
     An increase of
     Zs(blue) from 0.012 to 0.05 (comparable to that of the blue GCs in
     NGC 1399) changes substantially the appearance of that diagram  and
     makes the blue tilt (included in the model) less evident
     (right panel).\\

\noindent 2) The areal density distribution for blue and red clusters.

\begin{figure}
\resizebox{1.0\hsize}{!}{\includegraphics{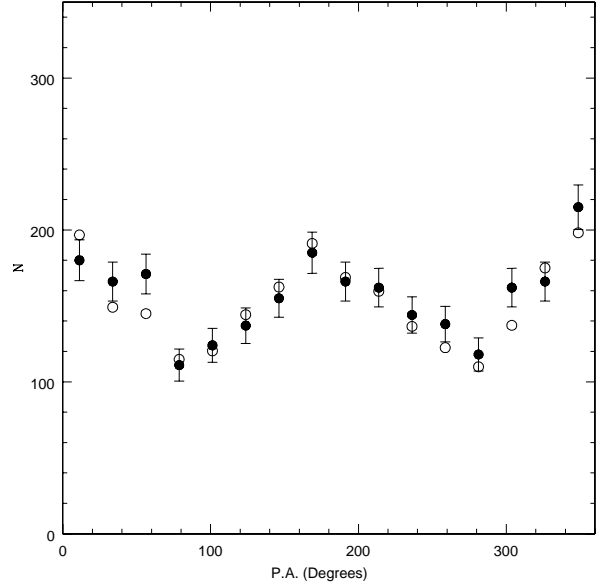}}
\caption{Azimuthal counts (within 22.5 degrees bins) for the globular clusters brighter than T$_1$=23.2 and (C-T$_1$) from 0.9 to 2.3 (large filled dots). The statistical uncertainty of the counts is shown with bars. Open dots come from a composite model that includes red clusters with a flattening q=0.8 and blue clusters with q=0.5 according to text.}
\label{Flat_4486}
\end{figure}

     The fact that the NGC 4486 GCS exhibits a noticeable flattening 
     has been pointed out by \citet{b45}. This feature is
     clearly seen in Figure \ref{Flat_4486},
     which shows azimuthal counts within a galactocentric circular 
     annulus with inner and outer radii of 120 and 360 arcsec, performed 
     using our photometry (T$_1$=21 to 23.2 and (C-T$_1$)=0.90 to 2.30).
     This figure also displays the results from a model that includes
     red GCs with a flattened spatial distribution with
     q=0.80, and blue clusters with q=0.50 (where q=b/a, is the ratio
     of the minor to major semi axis). The details of this model,
     that provides a consistent fit to the observations, are discussed
     below.

     The adopted flattenings come from the assumption that, if GCs
     trace a given stellar population, they  should share the same
     flattening. The inner  regions of NGC 4486 (where the red stellar
     component dominates the surface brightness) exhibits a q=0.85 (at 
     a=120 arcsec) and reaches about q=0.50 at the outermost detectable
     boundaries (see \citealt{b48}), where the blue stellar population should
     become more evident.

     The density run on a large angular scale was determined by using 
     Suprime camera observations by \citet{b81}. These
     authors determine areal density in circular annuli on a rectangular
     strip that extends to the east of the galaxy centre.

     We stress that, as those authors use the colour valley at (V-I)=1.10
     in their  photometry as a discriminant between both cluster 
     populations, their {\bf so called} blue clusters will eventually include
     the blue tail of the red population inferred from the colour histogram
     decomposition. We note that Tamura et al. also find that a projected
     NFW profile gives good representation of the areal density distribution 
     of the so defined blue GCs as FFG05 did in NGC 1399 using the
     same definition for the blue clusters.

     In order to test the compatibility of their observations
     with our approach, we generated a model that assumes that both
     cluster populations follow elliptical distributions with the
     flattenings mentioned before and a major axis coincident with
     that of the galaxy halo (P.A. $\approx$ 155 degrees).

     The density distribution of the genuine blue clusters in the inner
     regions of the galaxy was fit  using a surface density
     profile similar to that adopted for NGC 1399 but with a scale
     length rs=350 arcsec. This fit gives an adequate representation
     to the density depicted in Figure \ref{Pend_1399_4486} (lower panel).
   
     In turn, model red clusters were generated adopting a lowered Hubble
     density profile (or analytical King profile) with rc=60 arcsec 
     (from \citealt{b39}). In this case, the tidal radius 
     was  changed iteratively until the best fit to the Tamura et al.
     densities was obtained, yielding r$_t$=3600 arcsec.

     Model GCs colours  were generated as 
     described above while the T$_1$ magnitudes were derived
     by adopting  fully Gaussian integrated luminosity functions with
     turn-overs at T$_1$=22.9 and 23.2 and dispersions of 1.38 and 1.55 mags.
     for the blue and red globulars, respectively. These parameters were
     taken from \citet{b80}, who give values in the V band, and
     transformed to the T$_1$ band (through 
      $(V-R) = (V-T_1)=0.21(C-T_1)+0.19$).

     The completeness factors, that allow an estimate of the total number
     of GCs in each subpopulation from the fractional sampling in colour
     and magnitude, were 3.20 for the blue GCs and 3.31 for the red GCs. 
          
     The combined cluster population was then sampled in circular annuli,
     in order
     to compare with the Tamura et al. density profile and taking those GCs
     bluer than the colour valley at (C-T$_1$)=1.52  (or (V-I) $\approx$ 1.10,
     following their definition of the blue population). 

\begin{figure}
\resizebox{1.0\hsize}{!}{\includegraphics{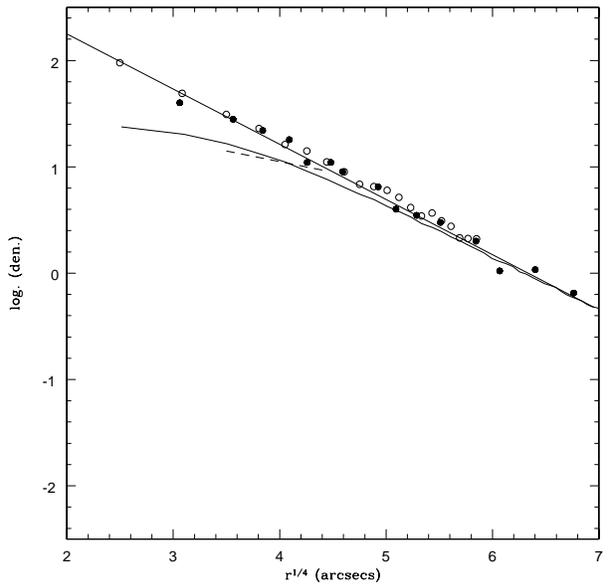}}
\caption{Projected areal distribution for NGC 4486 GCs bluer than the colour valley. Filled dots come from counts within circular annuli given by \citet{b81}. Open dots are from the model described in the text. The straight line is the NFW fit given by those authors. Note that, following this colour definition, the so called ``blue''  globulars do not exhibit the flat inner core. The adopted distribution of the ``genuine'' blue globulars is also shown (lower curve). The dashed line has the slope shown in figure \ref{Pend_1399_4486} for GCs bluer than (C-T$_1$)=1.21.}
\label{Dens_4486_B}
\end{figure}

     The result for {\bf this} blue population (shifted in log (dens.)) is 
     shown 
     in Figure \ref{Dens_4486_B} that also includes a straight line that 
     represents the NFW 
     profile (rs=226 arcsec; 20 kpc at their adopted galaxy distance) fit 
     by \citet{b80} to this cluster population. 
     The lower line in this diagram corresponds to the 
     genuine blue GCs and comes from sampling, also in circular annuli, the
     projection on the sky of an oblate ellipsoid (with q=0.5). This
     ellipsoid follows the blue GCs spatial density dependence mentioned above 
     with a cut off at 450 kpc from the galaxy centre.

     Figure \ref{Dens_4486_B} in fact shows that the model discussed in this section is
     able to match the Tamura et al. density profile, which does not show
     a flat core.

     The difference between this last profile and the adopted one
     for the genuine blue GCs, can thus
     be explained as the result of including the blue tail of the red 
     cluster population, characterised by a steeper spatial distribution,
     within the sample of GCs bluer than  (V-I)=1.10.
              
\begin{figure}
\resizebox{1.0\hsize}{!}{\includegraphics{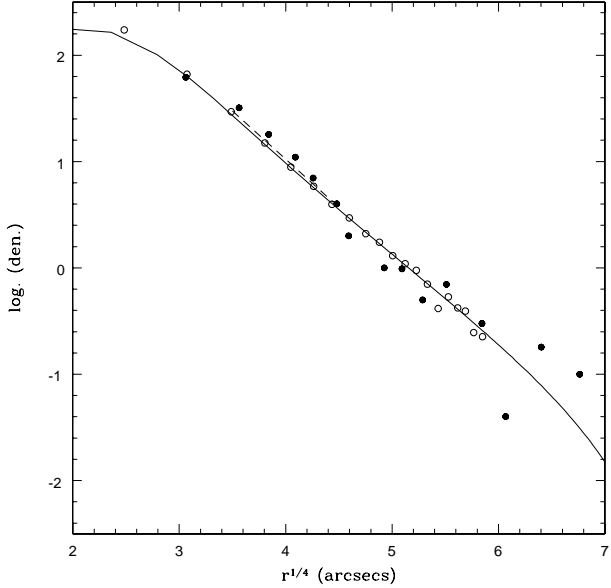}}
\caption{Projected areal distribution for NGC 4486 GCs redder than the colour valley (``red'' globulars). Filled dots come from counts within circular annuli given by \citet{b81}. Open dots come from the model described in the text. The model counts have been shifted vertically in order to take into account the limiting magnitude in that work. The dashed line has the slope shown in Figure 9 for
 GCs redder than C-T$_1$=1.72.}
\label{Dens_4486_R}
\end{figure}

     Figure \ref{Dens_4486_R}, in turn, shows the comparison of the model 
     with the
     red GCs as defined by Tamura et al. which, not suffering
     the colour overlapping effect, is directly comparable with our model
     red clusters.\\

\noindent 3) The surface brightness profile. 
 
     Two blue surface brightness profiles with a relatively large
     angular coverage are available for NGC 4486  in the literature:
     \citet{b10}  and \citet{b8}. These profiles,
     along the major axis of the galaxy, show good agreement up to
     a $\approx$ 600 arcsec where the Caon et al. profile becomes 
     systematically fainter. A comparison with the diffuse light
     map in the Virgo cluster by \citet{b48}, in turn,
     indicates a V surface brightness  V $\approx$ 26.5 mags. per
     arcsec$^{\sq}$ at a $\approx$ 1800 arcsec that imply B= 27.1 to
     27.5 mags. per arcsec$^{\sq}$ which is consistent with the Carter \&
     Dixon profile which we adopt in what follows.
     
\begin{figure}
\resizebox{1.0\hsize}{!}{\includegraphics{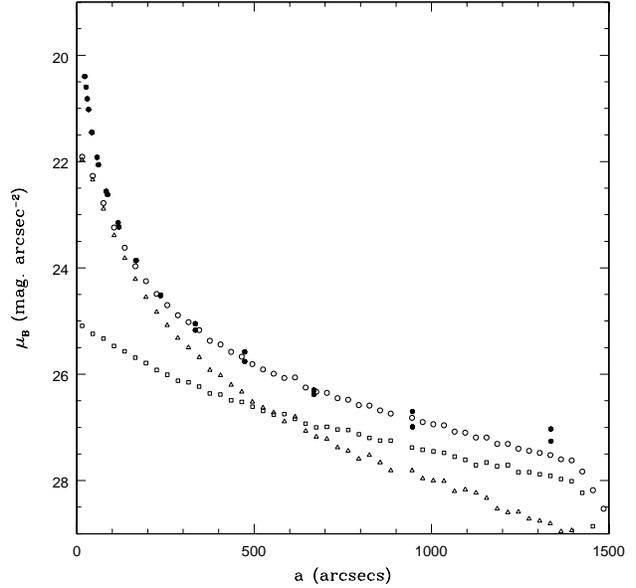}}
\caption{Observed B surface brightness profile for NGC 4486 ( \citealt{b10}; filled dots) confronted with the model fit (open circles). This last profile was obtained from a bi-dimensional model image processed with the ELLIPSE task within IRAF. Squares and triangles represent the luminosity associated with the ``blue'' and ``red'' stellar populations. Note that the model fails inside 100 arcsec in galactocentric radius where the globular distributions shows a flat core.}
\label{Prof_4486}
\end{figure}

     Figure \ref{Prof_4486} shows the best fit profile obtained through ELLIPSE
     from the blue synthetic image. The profile corresponds to a distance modulus
     (V-Mv)$_o$=31.0 and an interstellar reddening E(B-V)=0.022 from
     \citet{b70}.

     The profile fit requires $\gamma=1.18 (\pm 0.05) \times 10^{-8}$ and $\delta=1.2 \pm 0.1$
     with and yields an rms of $\pm$  0.07 mags.
     Again, and as already noted for NGC 1399, the flat core of the GCS
     does not allow a proper representation of the inner region of the
     galaxy. 

\begin{figure}
\resizebox{1.0\hsize}{!}{\includegraphics{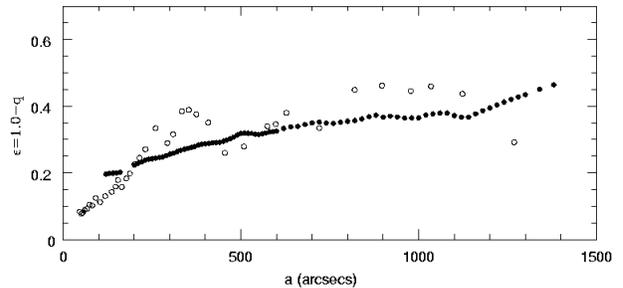}}
\caption{Ellipticity ($\epsilon=1-q$) variation of the NGC 4486 stellar halo as a function of semi-major axis $a$ from \citet{b10} (open circles), compared with the expected variation from the model fit (filled circles).}
\label{Ellip_4486}
\end{figure}

     The ouput from ELLIPSE shows that, as a result of composing two
     diffuse populations with different flattenings, the galaxy model
     flattening varies with galactocentric radius. This trend is
     compared with the ellipticity values ($\epsilon$=1-q) obtained by 
     \citet{b10} in Figure \ref{Ellip_4486}. The overall agreement is
     acceptable although it could be improved if the possibility
     of a variable q (for one or both cluster subpopulations) is allowed.
     However, the statistical uncertainties connected with the 
     azimuthal counts prevents a meaningful estimate of this eventual
     dependence.

\section{Dependence of Results on Uncertainties of the Fitting Parameters}
\label{DISC}

  The overall results from this modelling in terms of specific 
  frequencies, characteristic t* parameter (integrated over 
  metallicity for each of the cluster populations), diffuse
  stellar mass and $(M/L)_B$ ratios are listed in Table \ref{Model_values} .

\begin{table}
\centering
%\begin{minipage}
\caption{Model results for NGC 1399 and NGC 4486.}
\label{Model_values}
\begin{tabular}{@{}l@{}rrc@{}}
  &    NGC 1399        &     NGC 4486 &   \\
\hline
Adopted (V-Mv)o  &    31.4        &     31.0 &   \\
Zs(blue)         &  0.045     &    0.012      &  a\\
Zs(red)          &  1.45     &     0.90       &   \\
$\gamma$         &   0.82$\times 10^{-8}$   &  1.18$\times 10^{-8}$ &   \\
$\delta$         &   1.10      &      1.20 &   \\
number of blue globs. &     3900    &      7000     &   b\\
number of red globs.  &     4500    &      4800     &    b\\
$Sn^{*}$(blue globs)       &     12.1    &       27.9  &  c\\ 
$Sn^{*}$(red globs)        &    5.3     &        8.4   &  c\\
t* (blue globs)   &      3.44$\times 10^{-8}$ &   8.41$\times 10^{-8}$  &  d\\
t* (red globs)    &      0.85$\times 10^{-8}$ &   1.43$\times 10^{-8}$  &  d\\
t*  (total)       &      1.3$\times 10^{-8}$ &    2.82$\times 10^{-8}$ &  d \\
$(M/L)_B$ (blue pop)      &     4.2     &        3.9  &   \\       
$(M/L)_B$ (red pop)       &     9.6    &         8.7 &   \\
Total stellar mass ($M_{\odot}$)   &     7.2$\times 10^{11}$ &  4.8$\times 10^{11}$ &  e \\
Frac. mass (blue pop.)    &     0.18    &      0.20    \\
Frac. mass (red pop.)     &     0.82    &      0.80  &   \\
\hline
\multicolumn{4}{l}{\it \footnotesize a) Plus a blue ``tilt'': $\Delta Z=0.01(23.2-T_1)$}\\
\multicolumn{4}{l}{\it \footnotesize b) Inside a=1500 arcsec, assuming Gaussian LFs}\\
\multicolumn{4}{l}{\it \footnotesize c) Intrinsic values defined in terms of their associated stellar}\\ 
\multicolumn{4}{l}{\it \footnotesize  luminosities in the V band.}\\
\multicolumn{4}{l}{\it \footnotesize d) Integrated values defined in terms of their associated stellar masses.}\\
\multicolumn{4}{l}{\it \footnotesize e) Inside a projected galactocentric radius of 100 kpc}\\
\end{tabular}
%\end{minipage}
\end{table}

  The total stellar masses given in this table include a correction
  that takes into account the region within 120 arcsec in
  galactocentric radius, where the model does not provide an
  adequate fit. 
    
  In what follows we describe the uncertainties of these results
  in terms of the fitting parameters, $\gamma$ and $\delta$, as well
  as those connected with the colour-metallicity relation, (M/L) ratios,
  age, and adopted abundance scale.\\

\begin{figure}
\resizebox{1.0\hsize}{!}{\includegraphics{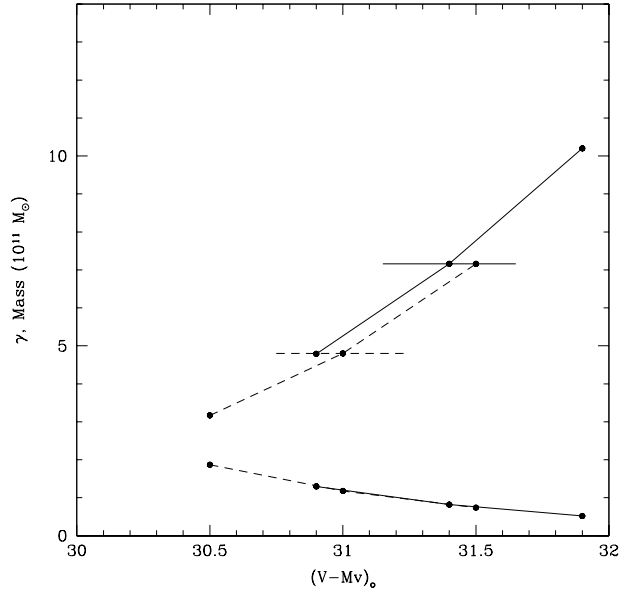}}
\caption{Variation of the $\gamma$ parameter (lower curves) and total projected stellar mass within 100 kpc (upper curves) obtained from model fits as a function of the adopted distance modulus (solid line: N1399; dashed: NGC 4486). Horizontal lines indicate a change of $\pm 0.25$ mags. around the adopted distance moduli.}
\label{gamma_var}
\end{figure}

\noindent -$\gamma$ parameter:         

  Figure \ref{gamma_var} depicts the dependence of both $\gamma$ and total 
  projected
  stellar mass (within a constant galactocentric radius of 100 kpc)
  with distance modulus. The
  adopted distance moduli for both galaxies are also shown along
  with a (formal) associated uncertainty of $\pm 0.25$ mags. Even
  such small uncertainties, do not rule out that both galaxies
  might be at  similar distances from the Sun, and in this case, the
  $\gamma$ parameter and total stellar masses of NGC 1399 and NGC
  4486 would also be very similar.\\

\noindent -$\delta$ parameter:
 
  $\delta$ is independent of the adopted distance and a variation
  of the order of the fit uncertainty ($\pm 0.1$), mainly impacts
  on the total mass  of the diffuse stellar population associated
  with the blue GCs, that changes by $\pm 15 \%$.
 
  Larger $\delta$ values imply a decrease in the mass of these stars
  and also redder integrated colours of the composite stellar
  population.\\

\noindent -M/L ratio and age:

  The (M/L) ratio depends on the age and metallicity
  of the seed GCs. We tentatively adopt an age of 12 Gy
  comparable to that of the Milky Way System (see, for example,
   \citealt{b13}).

  Synthetic population models by \citet{b83} show that a variation 
  of $\pm 2$ Gy around the adopted model age increases or decreases 
  those ratios by 15\% without changing the shape of the
  functional dependence with metallicity.  Accordingly,
  masses also change in the same proportion. Age variations of that order
  however, do not have a noticeable impact on either the shape of the
  brightness profile or its integrated colour.

  The characteristic $(M/L)_B$ ratios for both diffuse populations
  (as well as for the composite stellar population) were obtained
  by integrating over the whole range of abundances determined by
  their respective abundance scale lengths and adopting an upper
  cut off of $4 Z_\odot$. These results are not critically dependent upon 
  this formal upper limit.

   In particular, we note that the $(M/L)_B$ values of the red stellar
   populations are comparable to that obtained by 
   \citet{b66} ($(M/L)_B =10$ ) in the case of the central
   regions of NGC  1399 where the red population dominates the
   integrated luminosity.\\

\noindent -Abundance scale:

   The relation between [Fe/H]zw and [Z/H], adopted as a constant
   over the whole abundance range, might not be totally appropriate
   since the \citet{b46b} work does not include GCs at a 
   very low abundance regime. We also attempted models 
   adopting the Mendel et al. [Z/H]-[Fe/H]zw off set value (0.131) for the red GCs and
   a tentatively larger value (0.3) for the blue clusters. This
   modification leads to larger $(M/L)_B$ ratios for the blue
   population and to an increase of the total mass of about
   20\% (although no significant improvement of the fits of the colour
   histograms is obtained).

\section{Implications of the profile fit}
\label{IPF}

  The $\gamma$ and $\delta$ parameters that provide the best fit to the shape
  of the B band surface brightness profile of each galaxy  will also
  lead to a given:
\begin{enumerate}   
\item [a)] Galactocentric colour gradient of the galaxy halo.
\item [b)] Colour off-set between GCs and galaxy halo.
\item [c)] Behaviour of the cumulative GCs specific frequency with 
     galactocentric radius.
\item [d)] Metallicity distribution of the diffuse stellar population. 
\end{enumerate}
   A comparison of these predicted features with the observed ones, as
   follows, then may provide independent clues about the success of the
   model.\\

\noindent a) Galactocentric colour gradient of the galaxy halo.

   The main implication of the model is that, at galactocentric 
   distances larger than 120 arcsec, the main driver of the galaxy
   colour gradient is the (luminosity weighted) composition of the
   associated blue and red diffuse stellar populations. 

   FFG05 presented the expected colour (B-R) gradient for the 
   NGC 1399 halo on the basis that the colours of the diffuse
   stellar populations could be identified with the colours of
   the peaks of the associated GCs. That assumption is no longer necessary
   in this work as the colour of each mass element connected
   with a given globular, as well as its mass to luminosity ratio,
   are determined by the metallicity of the ``seed'' cluster.\\

\noindent b) Globulars-galaxy halo colour off set.

    The GCs {\bf mean} integrated colours (including all clusters) in
    massive elliptical galaxies  usually exhibit a galactocentric
    colour gradient comparable to that of the galaxy halo but blue ward
    shifted (\citealt{b78}; \citealt{b22}).

    In the context of the model discussed in this work, the cluster
    gradients arise as a consequence of (number weighting) averaging the two
    GCs subpopulations, characterised by different spatial scale lengths.
    The same reasoning apply 
    to the associated diffuse stellar populations but, in this case,
    weighted through the (M/L) ratios determined by metallicity, then
    leading to the observed colour off-set.

\begin{figure}
\resizebox{1.0\hsize}{!}{\includegraphics{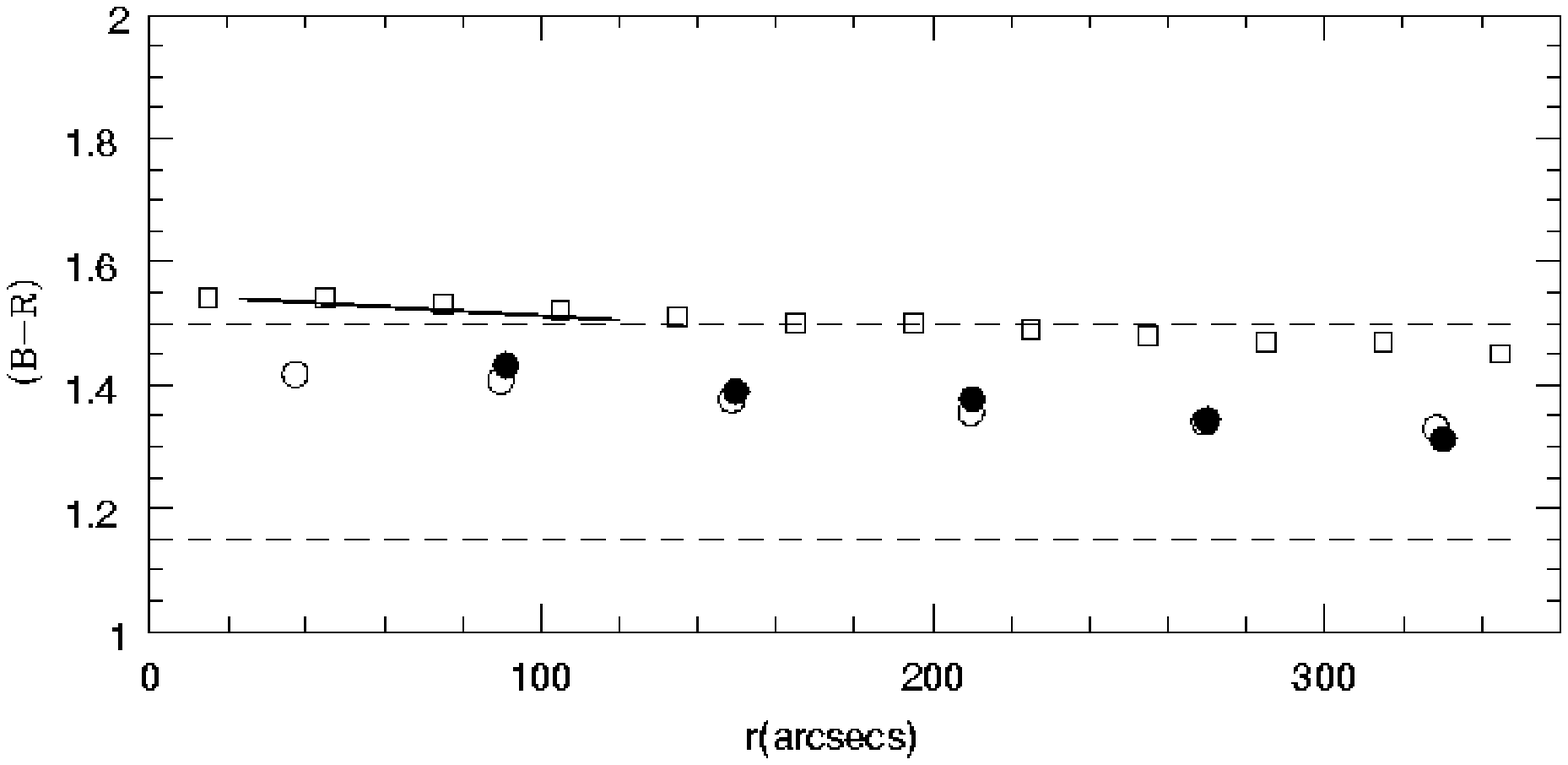}}
\resizebox{1.0\hsize}{!}{\includegraphics{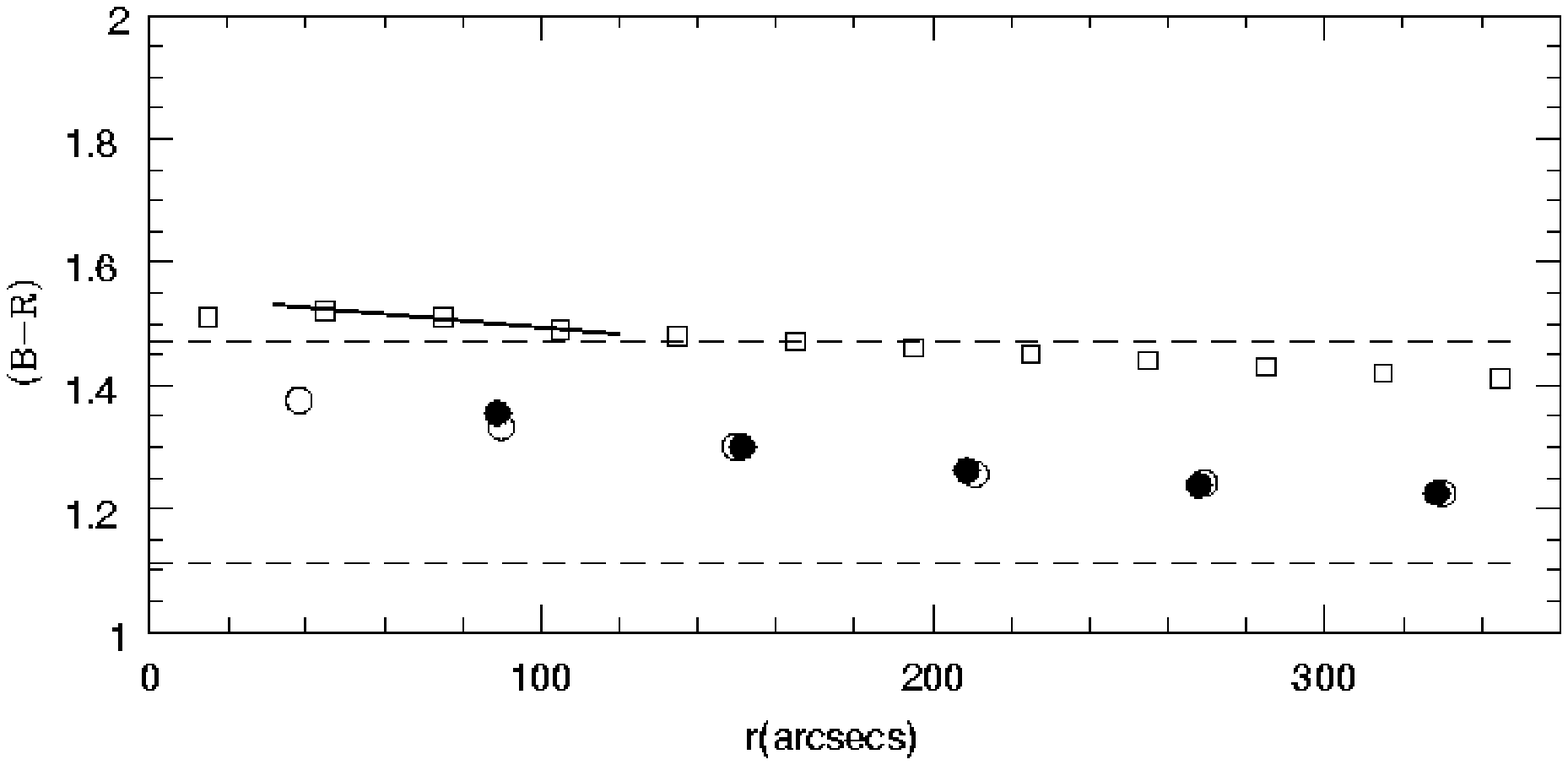}}
\caption{(B-R) colour gradient as a function of galactocentric distance for the halo (open squares) compared with the mean globular cluster colours from the model fit (open circles). The globular mean observed colours, derived from the photometry presented in this paper, are also shown (filled circles).  The short straight line indicates (B-R) colours derived from \citet{b47}. Upper panel: NGC 1399; Lower panel: NGC 4486. The dashed lines indicate, for each galaxy, the peak colours of the blue and red GCs.}
\label{colour}
\end{figure}

    Colour gradients derived from the profile fits are shown in Figure \ref{colour}. 
    In these diagrams, the predicted halo colours are compared
    with the mean globular integrated model colours and also with
    those obtained from the photometry presented in section
    II. These diagrams show that, in fact, the galaxy halos exhibit colour
    gradients comparable, but redder, than those of the GCs. 
    
     Figure \ref{colour} also shows that the (B-R) colour gradients determined
     by \citet{b47} in the inner regions of the galaxies are in very good
     agreement with with the predicted
     colours of the halo.\\

\noindent c) Cumulative globular cluster specific frequency.
 
\begin{figure}
\resizebox{1.0\hsize}{!}{\includegraphics{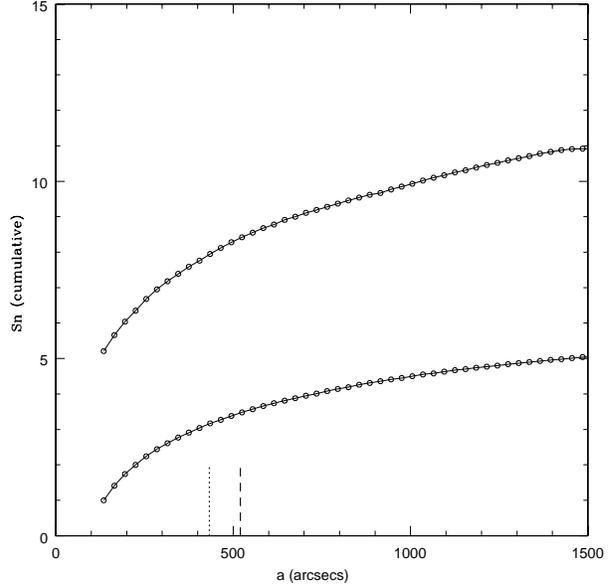}}
\caption{Cumulative globular cluster specific frequencies for NGC 1399 (lower curve) and NGC 4486 (upper curve) derived from the model fits. The vertical lines indicate galactocentric radius of 40 kpc (dotted line: NGC 1399; dashed line: NGC 4486) and indicate $Sn \approx 3.5$ and $Sn \approx 8.5$, respectively.} 
\label{Sn_cum}
\end{figure}

    Figure \ref{Sn_cum} shows the galactocentric variation of the 
    cumulative specific frequency derived from the best fit models. The GCs populations
    inside a galactocentric radius of 120 arcsec were taken from \citet{b17} and \citet{b39}.
    Even though each cluster subpopulation has its own intrinsic frequency, a variation of the
    {\bf composite} Sn is expected
    as the number ratio of blue to red GCs changes with galactocentric radius.

    The parametric Sn values \citep{b45}, defined at a
    galactocentric radius of 40 kpc, from this figure are $Sn \approx 3.5$ and 
    $Sn \approx 8.5$ for NGC 1399 and NGC 4486, respectively. These values 
    are considerably
    lower than previous estimates given in the literature as already noted
    in \citet{b24}.\\

\noindent d) The metallicity distribution of the diffuse stellar population.

    The shape of the  [Fe/H] distribution expected for the diffuse 
    stellar population was schematically derived in FFG05 on the
    basis of estimating a characteristic $(M/L)_B$ ratio and intrinsic
    specific frequency for each cluster population. In contrast, in this work
    each stellar mass element has a given metallicity,
    and hence (M/L) ratio. The statistic distribution of these masses,
    as a function of [Fe/H] is given in Figure 24 for
    NGC 1399 and NGC 4486. For both galaxies we show
    the inferred metallicity distribution, for galactocentric ranges from
    10 to 15  and 15 to 25 kpc convolved with a Gaussian kernel (dispersion: 0.20 dex)
    aimed introducing some degree of smoothing comparable to observational
    errors.

\begin{figure}
\resizebox{1.0\hsize}{!}{\includegraphics{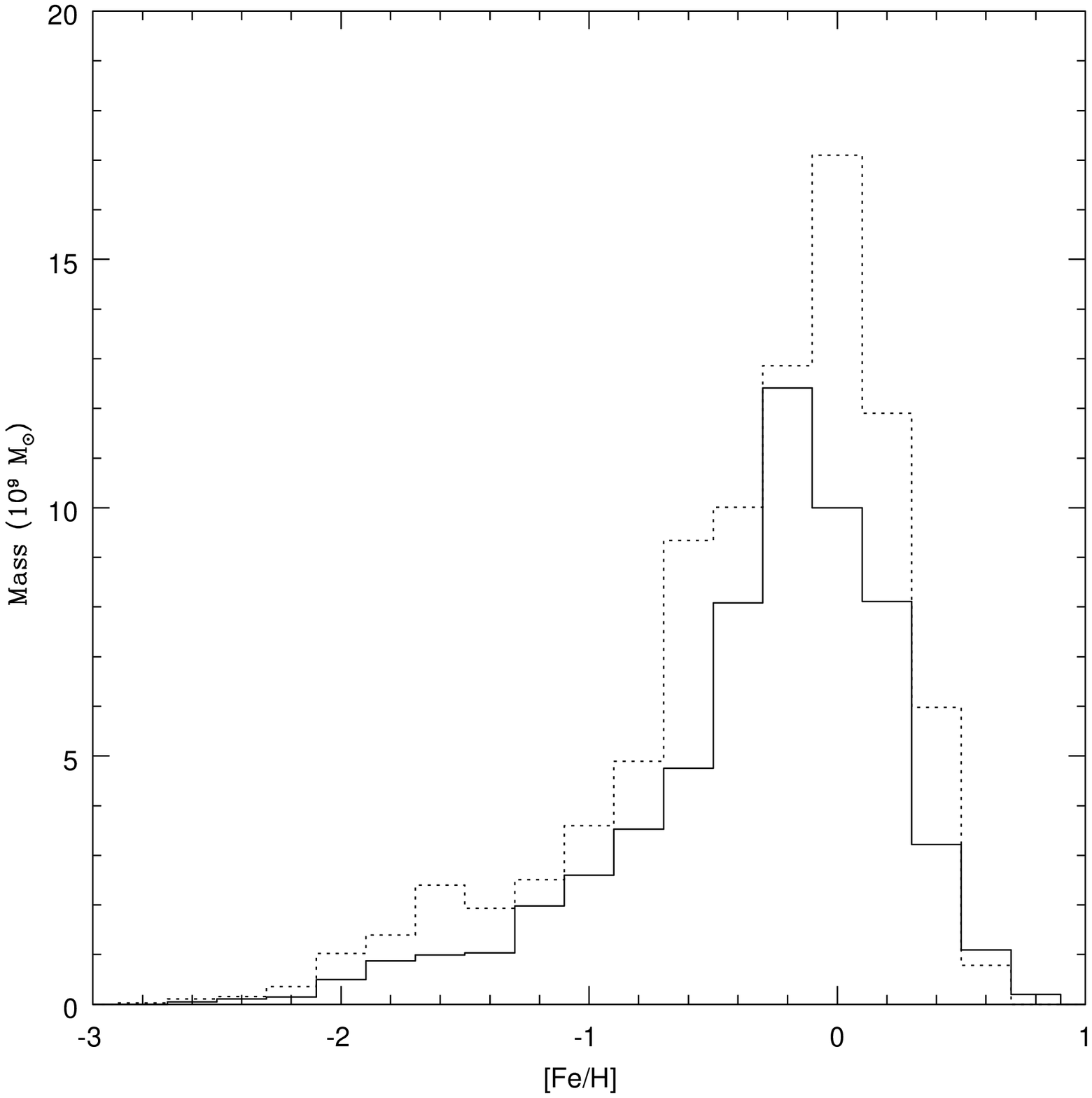}}
\resizebox{1.0\hsize}{!}{\includegraphics{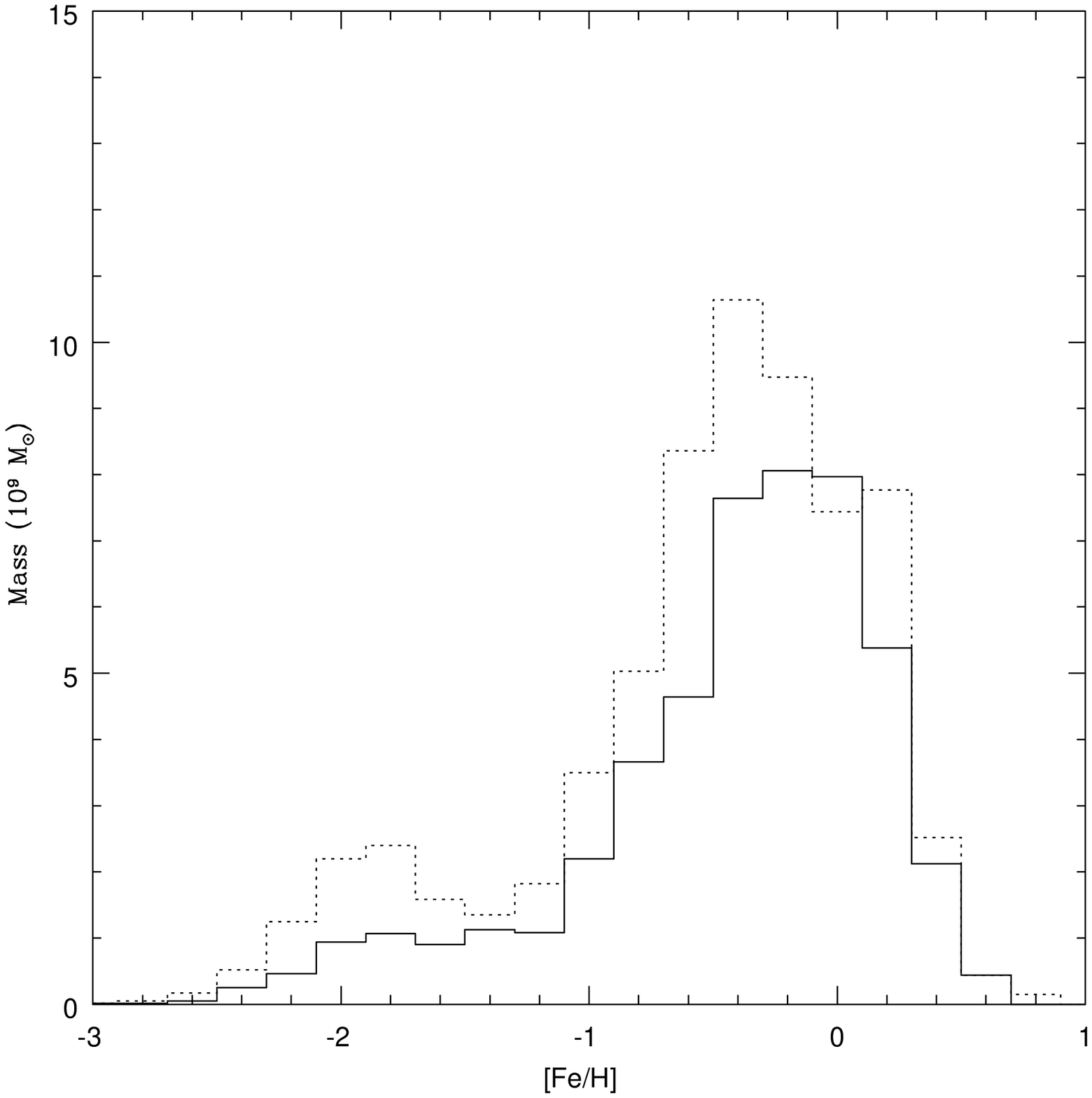}}
\caption{Stellar mass histogram as a function of metallicity for NGC 1399 (upper panel) and NGC 4486 (lower panel). These histograms belong to elliptical galactocentric radii between 10 and 15 kpc (solid line) and 15 to 25 kpc (dotted line) at an adopted distance of 19 Mpc (NGC 1399) and 15.8 Mpc (NGC 4486). The histograms asume an $\alpha$ ratio of 0.3} 
\label{Hist_Stellar}
\end{figure}

     These mass statistics can be transformed into star-number ones 
     under the assumption that the stellar luminosity functions do not
     depend strongly on metallicity.  
     A comparison with the cases of NGC 5128 (\citealt{b33} or \citealt{b62})
     and also M31 \citep{b15} 
     shows good qualitative agreement, i.e, the presence of a broad high
     metallicity component and an extended tail towards low metallicity
     that becomes more evident as galactocentric radius increases. 
     More recently, \citet{b50} presents stellar number statistics
     for a number of edge on spirals which also shows low metallicity skewed
     distributions, a feature that seems independent of the galaxy morphology.

\section{Caveats about the model}
\label{CAM}

 The model described in this work has several caveats, namely:\\

-colour bimodality is attributed to two different GCs subpopulations.
 An alternative view, based on the presence of an inflection region in
 the colour-metallicity relation as the main driver of the shape of the
 colour histograms, has been suggested by \citet{b84}.
  So far, however, neither our
 calibration nor recent spectroscopic results in NGC 4472 \citep{b76}
 seem  to support that situation. Further results in this last 
 direction can also be found in \citet{b40} for the case of NGC
  4486.

  - Although an exponential dependence of the number of GCs
   with abundance was adopted,  more complex functional 
  dependences, hidden by the noise of the GCs statistics, could not be
 ruled out.

 - An abundance dependence with brightness, that leads to a blue tilt
   in the case of the blue GCs, is included in the models.
 However, we cannot infer whether the tilt is just a local effect or
 if it is  actually shared by the diffuse
 stellar population associated with the blue clusters. 
 Nevertheless removing
 the tilt from the models, has little impact on the output integrated colours
 that then would become slightly bluer ($\approx 0.015$ mags. in (B-R)).

  -The two dominant cluster subpopulations are assumed to be coeval.
  Further refinement of this aspect could be incorporated once 
  meaningful ages become available. On one side some works (e.g
  \citealt{b37}) do not find detectable age differences between
  the blue and red populations in NGC 4486. On the other, (e.g.
  \citealt{b19}; \citealt{b55}; \citealt{b56}, or \citealt{b35}) do 
  find a certain fraction of intermediate age clusters in NGC 1399
  and in other galaxies.

  -The $[\alpha/ Fe]$ is adopted as  constant and equal for the
  cluster subpopulations in both galaxies in order to derive the
  [Fe/H] stellar distributions . Even though
  the adopted value is appropriate for the MW (see \citealt{b6}, or
  \citealt{b82}) and also representative for 
  ellipticals \citep{b58}, a possible variation with 
  metallicity  as noted by these last authors and, earlier, by \citet*{b69},
  cannot be ruled out.

\section{Discusssion and Conclusions}
\label{Conclu}

  The results presented in previous sections show that the surface
 brightness profiles of both NGC 1399 and NGC 4486 can be traced using a 
 common link between GCs and the stellar halo populations in these 
 galaxies. This link imply that the number of GCs
 per diffuse stellar mass, defined as {\bf $t=\gamma \exp(-\delta[Z/H])$ },increases 
 when chemical abundance decreases. We note 
 that \citet{b33} had already found an increase of Sn with decreasing 
 GCs metallicity in the case of NGC 5128.

 This suggests that, on  a large scale, the dominant globular cluster subpopulations
 formed along major star forming episodes and following a similar
 pattern. However, it is not yet clear whether abundance, through the role that
 it plays in the t parameter, is the physical reason that governs the 
 fraction of clustered to diffuse stellar mass or, eventually,
 some other ``hidden'' variable in turn correlated with abundance.

 The quality of the profile fits is comparable to any other parametric
 approximation in a range that covers  a galactocentric radius
 from 10 to 100 kpc. In the inner regions, GCs fail to map the brightness
 distribution probably as a consequence of cluster destruction processes due
 to gravitational effects. However, it seems that survivor clusters, with
 large perigalacticon orbits, are still able to trace their associated stellar
 populations. It is worth mentioning that, based on far UV observations of 
 NGC 1399 , \citet{b42}, find arguments that support the coexistence
 of two widely different stellar populations, in terms of chemical abundance,
 in the nucleus of the galaxy.

 Even though the approach only aims at reproducing the brightness profiles,
 other connected features as galactocentric colour gradients, GCs-halo
 colour offset, cumulative cluster specific frequency and inferred stellar
 metallicity distributions compare very well with observations.

 It seems also remarkable that the common quantitative GCs-stellar halo link
 works in both galaxies although their cluster populations exhibit detectable
 differences in cluster numbers and chemical abundance. Both GCs subpopulations have
 larger abundance scale lengths (as defined in Section 4) in NGC 1399 than in NGC
 4486 leading to total mean abundance ratios
 of 1.4 and 2.5 for the red and blue GCs, respectively.

 As shown in Figures 10 and 14, the approach presented in this paper delivers GCs
 colour distributions that are not strongly different  from a two Gaussians fit
 requiring five free parameters, a common  procedure in the literature (e.g Ostrov,
 Forte \& Geisler 1998). However, we note that our models indicate a lower ratio
 of the number of blue to red GCs and only require three free parameters.

 Blue GCs in NGC 4486 have a  small abundance scale lenght, about four times
 smaller than that of the blue GCs in NGC 1399, and we speculate that this may
 be connected with a shorter formation time scale. In turn, that lower abundance
 spread may be the reason behind the presence of a blue tilt, connected with
 cluster mass, in the colour magnitude diagram. This feature seems absent,
 or probably masked, by the larger abundance scale of the blue GCs in NGC 1399,
 a situation that may also hold in other galaxies (e.g. NGC 4472, Strader et al. 2006).

 This last result argues in favour of the idea that blue GCs ``know'' about the
 galaxy they are associated with (\citealt*{b77b}). In any case, and
 in both galaxies, the blue GCs barely reach an abundance close
 to [Z/H]=-0.5. The
 reason for this upper metallicity cut off may be connected with some kind of sincrhonising
 event as the re-ionisation of the Universe
 (\citealt{b11}; \citealt{b67}; \citealt{b63}). However, the different 
 abundance scale lengths of the blue GCs also suggest that a local phenomenon,
 distinct for each galaxy, may have also played a role in modulating the
 the star formation rate (e.g. the onset of galaxy nuclear activity).

 The overall picture seems consistent with some scenarios already
 discussed in the
 literature (\citealt{b18}; \citealt{b34}) that invoke two different cluster
 formation mechanisms and, probably, environmental conditions. 

 In contrast with the relative abundance homogeneity and large
 spatial (half density) core radii ($\approx$ 25 kpc) of the blue
 GCs, the red ones exhibit a large abundance heterogeneity and
 much smaller core radii ($\approx$ 5 kpc) also shared by the red diffuse stellar
 population. This degree of heteregeneity may be connected , for example, to mergers of different
 nature (e.g. \citealt{b72}).

 The total globular cluster formation efficiencies, in terms of stellar
 mass, indicated by the models, and adopting an average cluster mass
 of $2.5 \times 10^{5 } M_\odot $, are $2.4 \times 10^{-3}$ for NGC 1399 and
 $5.6 \times  10^{-3}$ for NGC 4486. They are comparable to the efficiency 
 derived by \citet{b46} although, in that case, the definition of 
 efficiency included total baryonic mass.
       
 Blue clusters  show a higher formation efficiency (in terms
 of the stellar mass they are associated with), when compared with the
 red ones, probably as a consequence of a lower star formation efficiency
 during the early phases of galaxy formation at a low metallicity regime. 

 Although there is no strong evidence of an age difference between the
 blue and red globular subpopulations (e.g. \citealt{b37}) within the
 uncertainties of the measurements, a possible temporal sequence
 that assumes
 the formation of the blue population first, cannot be discarded as
 a result of the relatively small time scales involved 
 at the early phases of galaxy formation (see \citealt{b3b}). In this frame,
 the chemical enrichment provided by a presumably progenitor blue population 
 might be important to boost later stellar formation efficiency through
 an abundance enrichment that may reach $[Z/H] \approx $-0.60 within 100 kpc
 of the galaxy nucleus.

 Some  scenarios suggest that blue GCs formation is associated with
 dark matter (\citealt{b3}; \citealt{b49}; \citealt{b57}), and it
 is tempting to look for such a connection. For example, the
 results listed in Table \ref{Model_values} show that while both
 galaxies have a similar total number of red GCs, NGC 4486
 outnumbers NGC 1399
 in a factor of about 1.8 in terms of blue GCs. Dark mass
 estimates within a galactocentric radius of 100 kpc are
 3.4$\times 10^{12} M_\odot$ for NGC 1399 (extrapolating
 data from \citealt{b64})
 and 7.4$\times 10^{12} M_\odot$ for NGC 4486 \citep{b10b},
 leading to a ratio $\approx 2.0$ comparable to that in the number
 of blue GCs.

 The similarity of the stellar galaxy masses,
 and the difference in their total masses, had already been noticed by
 \citet{b36} on the basis of their X ray analysis.
   
 FFG05  found that, adopting {\bf their} definition of blue clusters, the 
 density profiles of the NGC 1399 GCs could be fit with a NFW profile
 \citep{b51} with a scale length of 375 arcsec,
 coincident with that derived for the inferred dark matter halo by
 \citet{b64}. \citet{b80} also perform  a NFW profile fit to 
 the blue clusters in NGC 4486. However, both approaches deserve a revision
 since, on the basis of the results presented in this work, the
 ``genuine'' blue GCs exhibit a rather extended inner core in their surface
 density profiles. As shown here, these cores have been disguised by the
 inclusion of the blue tail of the red subpopulation within the ``blue'' GCs
 sample. This overlapping should be even more severe when using colour
 indices less sensitive to metallicity than (C-T$_1$) and should be taken
 into account when doing, for example, kinematic analysis of the cluster
 subpopulations. \\

\section*{Acknowledgments}
     This work was supported by grants from La Plata National University,
     Agencia Nacional de Promocion Cientifica y Tecnologica, and CONICET,
     Argentina. DG gratefully acknowledges support from the
     Chilean {\it Centro de Astrof\'isica } FUNDAP No 15010003.

\label{lastpage}
\end{document}